\documentclass[prd,onecolumn]{revtex4}
\usepackage{dcolumn}
\usepackage{multirow}
\usepackage{graphicx}
\usepackage{amssymb}
\usepackage{bm}
\usepackage{hyperref}
\usepackage{epstopdf}
\usepackage{color}
\usepackage{mathrsfs}
\usepackage{amsmath,amssymb,amsthm}
\begin{document}
\title{Exploring the expansion dynamics of the universe from galaxy cluster surveys}
\author{Deng Wang}
\email{Cstar@mail.nankai.edu.cn}
\affiliation{Theoretical Physics Division, Chern Institute of Mathematics, Nankai University,
Tianjin 300071, China}
\author{Xin-He Meng}
\email{xhm@nankai.edu.cn}
\affiliation{Department of Physics, Nankai University, Tianjin 300071, China}
\begin{abstract}
To understand the expansion dynamics of the universe from galaxy cluster scales, using the angular diameter distance (ADD) data from two different galaxy cluster surveys, we constrain four cosmological models to explore the underlying value of $H_0$ and employ the model-independent Gaussian Processes to investigate the evolution of the equation of state of dark energy. The ADD data in the X-ray bands consists of two samples covering the redshift ranges [0.023, 0.784] and [0.14, 0.89], respectively. We find that: (i) For these two samples, the obtained values of $H_0$ are more consistent with the recent local observation by Riess et al. than the global measurement by the Plank Collaboration, and the $\Lambda$CDM model is still preferred utilizing the information criterions; (ii) For the first sample, there is no evidence of dynamical dark energy (DDE) at the $2\sigma$ confidence level (CL); (iii) For the second one, the reconstructed equation of state of dark energy exhibits a phantom-crossing behavior in the relatively low redshift range over the $2\sigma$ CL, which gives a hint that the late-time universe may be actually dominated by the DDE from galaxy cluster scales; (iv) By adding a combination of Type Ia Supernovae, cosmic chronometers and Planck-2015 shift parameter and HII galaxy measurements into both ADD samples, the DDE exists evidently over the $2\sigma$ CL.

\end{abstract}
\maketitle

\section{Introduction}
At the end of twentieth century, two cosmological groups discovered that the universe is undergoing a phase of accelerated expansion by using the Type Ia supernovae (SNe Ia) observations \cite{1,2}. In the past about two decades, this mysterious phenomenon is well confirmed by the most recent SNe Ia data \cite{3} and other astronomical observations such as cosmic microwave background (CMB) radiation \cite{4}, baryonic acoustic oscillations (BAO)\cite{5}, observational Hubble parameter \cite{6}, and so forth. To explain the accelerated mechanism, cosmologists have proposed an exotic and negative pressure fluid dubbed dark energy (DE). To date, the realistic nature of DE is still unknown, but its main properties are substantially explicit: (i) it must be homogeneously and isotropically distributed on large cosmological scales; (ii) the modulus of its effective pressure $p$ needs to be comparable to its energy density, namely $|p|\sim\rho$. In the literature, the simplest model to depict DE is the concordance model of cosmology, i.e., the so-called $\Lambda$-cold-dark-matter ($\Lambda$CDM) model, which is characterized by the equation of state (EoS) of DE $\omega=-1$. Although the $\Lambda$CDM model can explain successfully many aspects of the observational universe, it still faces two fatal problems, namely the fine-tuning and coincidence problems \cite{7}. The former implies that the theoretical value for the vacuum energy density are far larger than its observational value, i.e., the well-known 120-orders-of-magnitude discrepancy that makes the vacuum explanation very puzzling; while the latter indicates why the energy densities of the dark matter (DM) and DE are of the same order at the present time, since their energy densities are so different from each other during the evolutional process of the universe. Recently, using new calibration techniques and indicators, Riess et al. reported the improved local measurement of the Hubble constant $H_0=73.24\pm1.74$ km s$^{-1}$ Mpc$^{-1}$ (hereafter R16) \cite{8}, which exhibits a stronger tension with the Planck 2015 release $H_0=66.93\pm0.62$ km s$^{-1}$ Mpc$^{-1}$ (hereafter P15) \cite{9} at the 3.4$\sigma$ CL. All these facts suggests that the true nature of DE may not be the cosmological constant $\Lambda$, and pose two great challenges to explore the expansion dynamics for cosmologists:

$\star$ Is actually DE a time-dependent physical component ($\omega\neq-1$) or dominated by a cosmological $\Lambda$ term ?

$\star$ Theoretically, new physics is urgent to be mined to explain the current $H_0$ tension; Experimentally, how to determine more reasonably the value of $H_0$ with higher accuracy ?

To address these two issues, using the model-independent Gaussian Processes (GP), we have performed the improved constraints on the EoS of DE in light of recent cosmological data including 580 SNe Ia, 30 cosmic chronometers and Planck-2015 shift parameter \cite{10}, which indicates that the $\Lambda$CDM model is still supported by these data and the results of reconstructions support substantially R16's local measurement of $H_0$. Then, we also explore the values of $H_0$ and constrain the EoS of DE by only using the latest HII galaxy measurements \cite{11}, and find that the obtained values of $H_0$ are more consistent with the R16's local observation than P15's global measurement, and the $\Lambda$CDM model can fit the data well at the $2\sigma$ CL. Following this logical line, we are full of interest in exploring the underlying value of $H_0$ in a larger local scale than HII galaxies. As a consequence, we continue investigating the expansion dynamics of our universe from galaxy cluster scales. As is well known, galaxy clusters are the largest gravitationally collapsed structures in the universe, with a hot diffuse plasma ($T\sim10^7-10^8 K$) that fills the intergalactic space, and they are also important cosmological probes to distinguish various cosmic evolutional models \cite{12}.

This paper is organized as follows: In Section 2, we describe the ADD data used in this analysis. In Section 3, we constrain four cosmological models by using the ADD data. In Section 4, we employ the GP method to constrain the EoS of DE. The discussions and conclusions are presented
in the final section.

\section{The ADD data}
In this analysis, we adopt two galaxy cluster samples, which are based on different morphologies and dynamics, to explore the expansion dynamics of the universe. These two samples has been widely used to test the validity of the Einstein equivalence principle combining with other cosmological probes such as SNe Ia and strong gravitational lensing \cite{r1,r2,r3}.

The first sample consists of 25 galaxy clusters lying in the redshift range $z\in [0.023, 0.784]$ from \cite{12}. Motivated by images from the Chandra and XMM-Newton telescopes, which shows an elliptical surface brightness of galaxy clusters, the authors utilized an isothermal elliptical $\beta$ model to depict the galaxy clusters, and constrain the intrinsic shapes of galaxy clusters to obtain the ADD data by combining X-ray and Sunyaev-Zel'dovich (SZ) observations. The 25 ADD data points were obtained for two sub-samples: 18 galaxy clusters from \cite{13} and 7 from \cite{14}, where a spherical $\beta$ model was assumed.

The second sample are formed by 38 galaxy clusters in the redshift range $z\in [0.14, 0.89]$ assuming the hydrostatic equilibrium model, which were obtained by using X-ray data from Chandra and SZ effect data from the Owens Valley Radio Observatory and the Berkeley-Illinois-Maryland Association interferometric arrays \cite{15}. It is worth noting that, assuming generalized $\beta$ spherical models, the authors obtained the ADD data by analyzing the cluster plasma and dark matter distributions. As described in \cite{15}, all the data points are almost followed by asymmetric uncertainties. To deal with this, we adopt a simple dealing method to obtain the data with symmetric uncertainties. In \cite{16}, this method has been used to acquire data for comparing different morphological models of galaxy clusters (i.e., elliptical $\beta$ model and spherical $\beta$ model) through model-independent tests of cosmic distance duality relation (CDDR), and it can be concluded as
\begin{equation}
E(D_A)=D_A, \qquad \sigma_{D_A}=\mathrm{max}(\sigma_+,\sigma_-) \label{1},
\end{equation}
where $D_A$, $E(D_A)$, $\sigma_{D_A}$, $\sigma_+$ and $\sigma_-$ denote the ADD, expected value of ADD, $1\sigma$ standard deviation of ADD, the upper and lower limits of data error, respectively. More specifically, we use the reported value of ADD $D_A$ as the expected value $E(D_A)$ and the larger flank of each two-sided error as the $1\sigma$ standard deviation $\sigma_{D_A}$.

In history, Etherington verified the CDDR based on the following two assumptions for the first time in 1933 \cite{17}:

$\star$ The light travels always along the null geodesics in a Riemannian geometry;

$\star$ The number of photons is conserved over during the evolutional process of the universe.

The CDDR is also called Etherington's reciprocity relation, and it connects two different scale distances via the identity
\begin{equation}
\frac{D_L}{D_A(1+z)^2}=\eta=1 \label{2},
\end{equation}
which relates the luminosity distance (LD) $D_L$ and the ADD $D_A$ at the same redshift $z$. It is noteworthy that, using the current astronomical observations, one can test the correctness of the general metric theories of gravity including the Einstein's one, which correspond to the case of $\eta=1$ (for details, see \cite{18}). In a Friedmann-Robertson-Walker (FRW) universe,  the expression of the LD $D_L(z)$ can be written as
\begin{equation}
D_L(z)=\frac{1+z}{H_0\sqrt{|\Omega_{k}|}}\mathrm{sinn}\left(\sqrt{|\Omega_{k}|}\int^{z}_{0}\frac{dz'}{E(z';\theta)}\right), \label{3}
\end{equation}
where $\theta$ denotes the model parameters, $H_0$ is the Hubble constant, the dimensionless Hubble parameter $E(z; \theta)=H(z; \theta)/H_0$, the present-day cosmic curvature $\Omega_{k}=-K/(a_0H_0^2)$,  and for $\mathrm{sinn}(x)= \mathrm{sin}(x), x, \mathrm{sinh}(x)$, $K=1, 0, -1$ , which corresponds to a closed, flat and open universe, respectively.

In our analysis, we transform the ADD data to the available effective LD data at the same $z$ by assuming $\eta=1$. Then, in order to constrain different cosmological models, we perform the so-called $\chi^2$ statistics using different expressions of the LD:
\begin{equation}
\chi^2=\sum^{N}_{i=1}[\frac{D_{L_{obs}}(z_i)-D_{L_{th}}(z_i;\theta)}{\sigma_i}]^2, \label{4}
\end{equation}
where $\sigma_i$, $D_{L_{obs}}(z_i)$ and $D_{L_{th}}(z_i)$ denote the $1\sigma$ error, the effectively observed and theoretical value of the LD at a given redshift $z_i$ for every galaxy cluster, respectively, and $N$ denotes two different sample sizes (i.e., for the first and second samples, N=25 and 38, respectively).

\section{The constraints on DE models }
In order to investigate the values of $H_0$, we constrain four different cosmological models by using the transformed LD data from galaxy clusters.
These four models are, respectively, the spatially flat $\Lambda$CDM, non-flat $\Lambda$CDM, $\omega$CDM and decaying vacuum (DV) models.

The dimensionless Hubble parameter for the spatially flat $\Lambda$CDM model ($\omega=-1$) is
\begin{equation}
E(z)=\sqrt{1-\Omega_{m}+\Omega_{m}(1+z)^3}, \label{5}
\end{equation}
and for the spatially non-flat $\Lambda$CDM model it is written as
\begin{equation}
E(z)=\sqrt{1-\Omega_{m}+\Omega_{k}(1+z)^2+\Omega_{m}(1+z)^3}, \label{6}
\end{equation}
where $\Omega_{m}$ is the dimensionless matter density ratio parameter at the present epoch.

In the spatially-flat $\omega$CDM parametrization we have
\begin{equation}
E(z)=\sqrt{(1-\Omega_{m})(1+z)^{3(1+\omega)}+\Omega_{m}(1+z)^3}, \label{7}
\end{equation}
where $\omega$ is the constant, negative, EoS parameter connecting the DE fluid pressure with energy density through $p=\omega\rho$.

Another consideration is the so-called DV model, which is aimed at resolving the famous fine-tuning problem by assuming the cosmological constant to be dynamical. Generally, to obtain a definite DV model, one should specify a vacuum decay law. Nonetheless, Wang and Meng proposed an interesting model based on a simple assumption about the form of the modified matter expansion rate \cite{19}. The corresponding dimensionless Hubble parameter for this DV model is expressed as
\begin{equation}
E(z)=\sqrt{1-\frac{3\Omega_{m}}{3-\epsilon}+\frac{3\Omega_{m}}{3-\epsilon}(1+z)^{3-\epsilon}}, \label{8}
\end{equation}
where $\epsilon$ is a small positive constant characterizing the deviation from the standard matter expansion rate.
\begin{figure}
\centering
\includegraphics[scale=0.4]{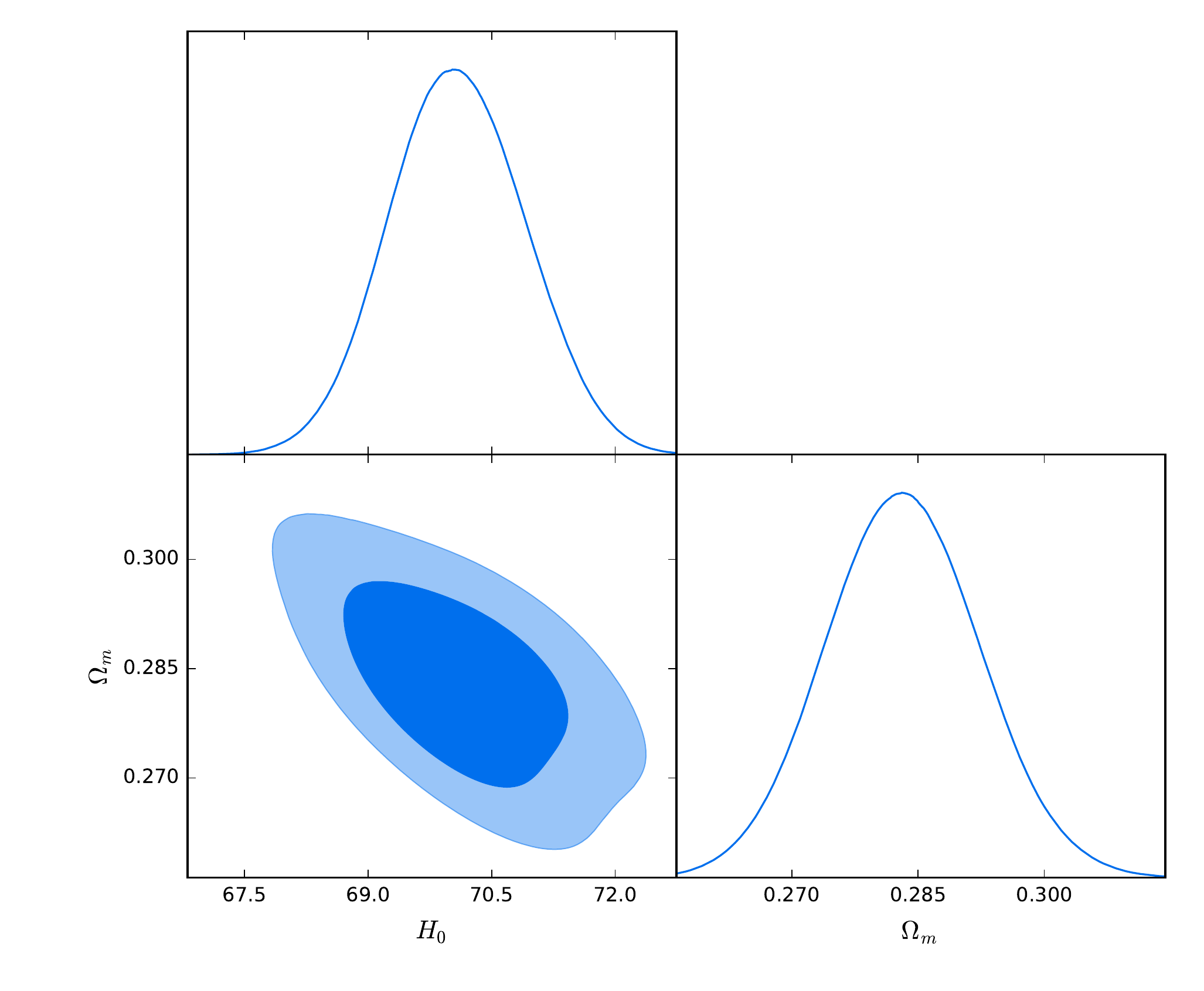}
\includegraphics[scale=0.4]{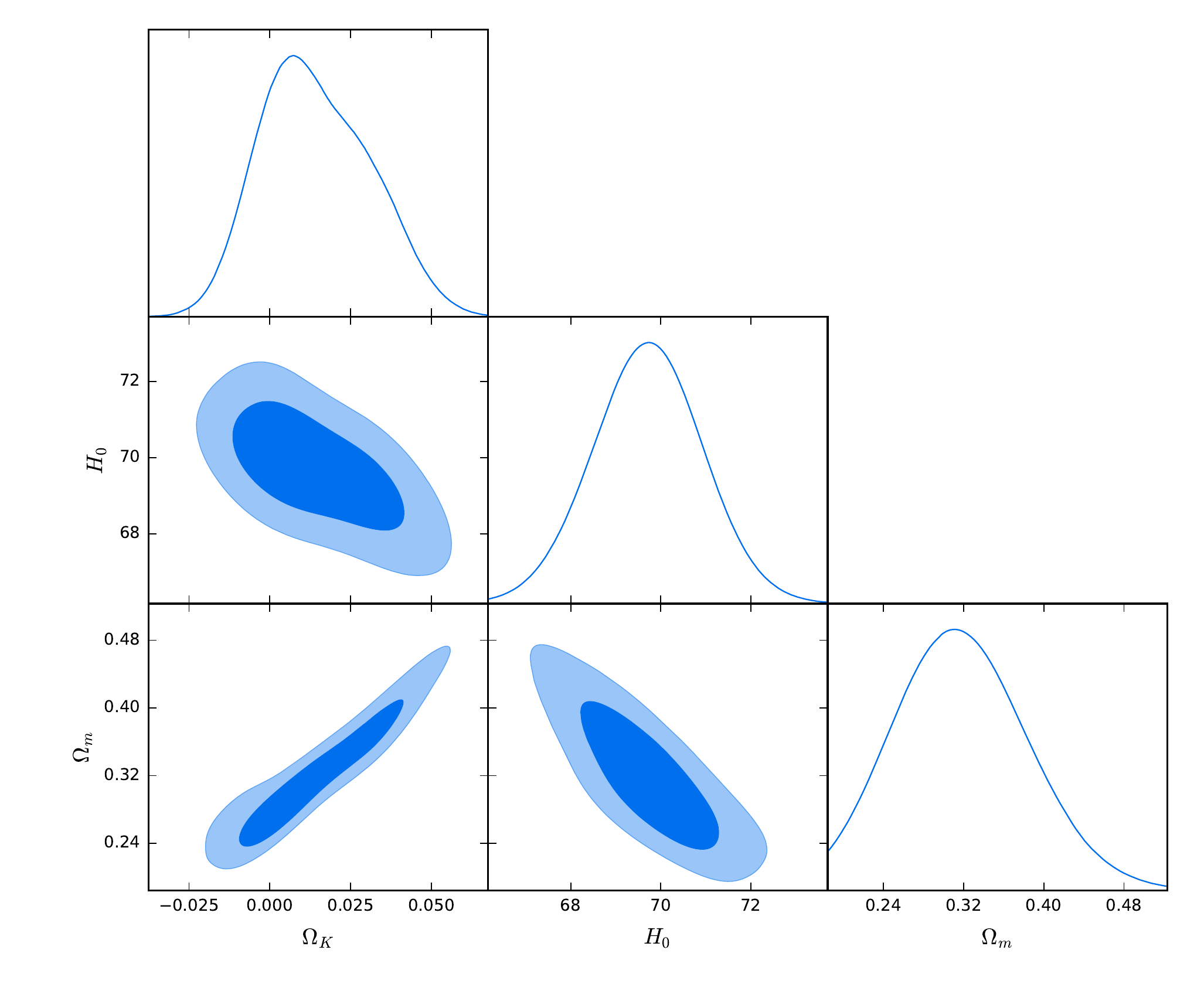}
\includegraphics[scale=0.4]{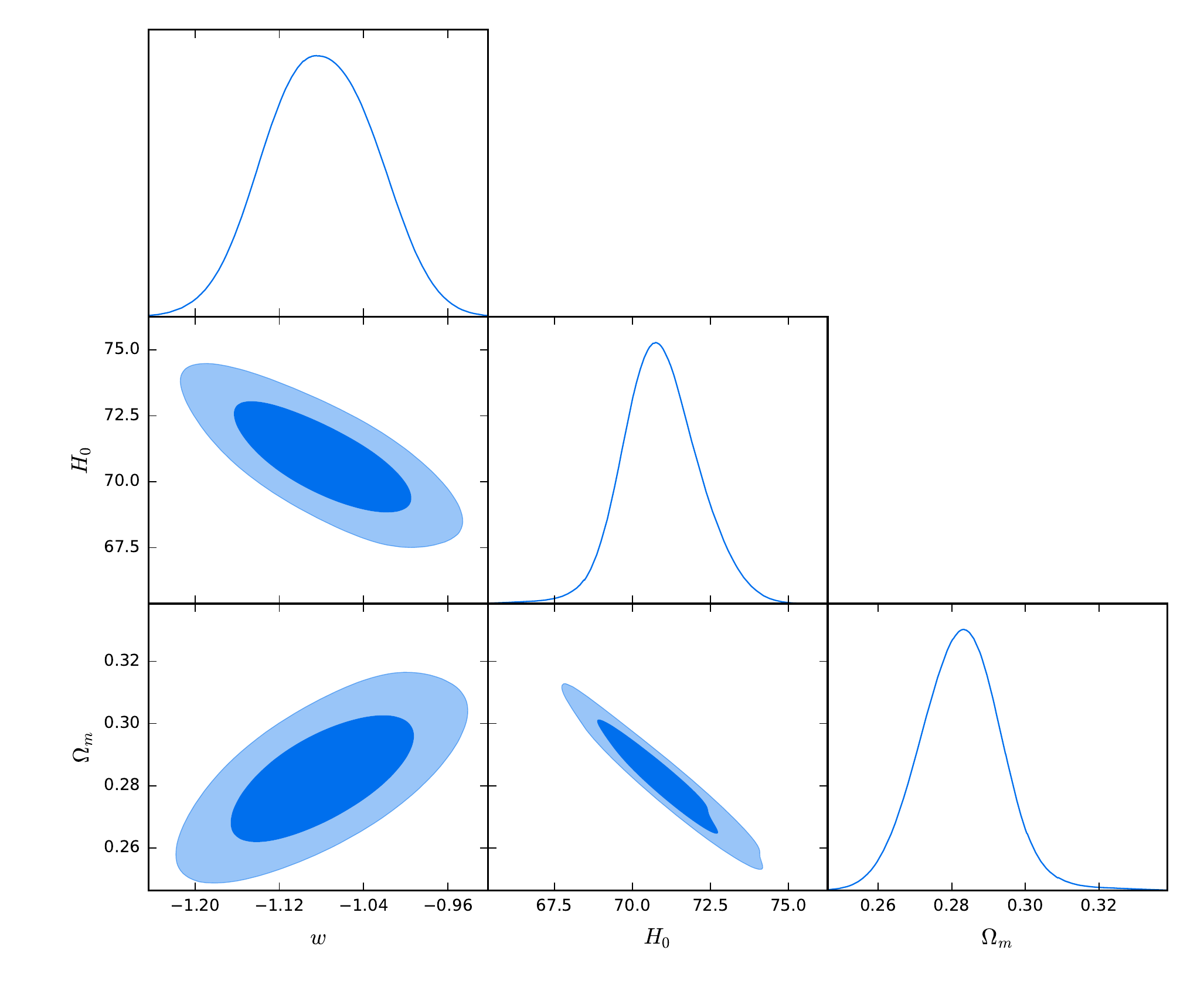}
\includegraphics[scale=0.4]{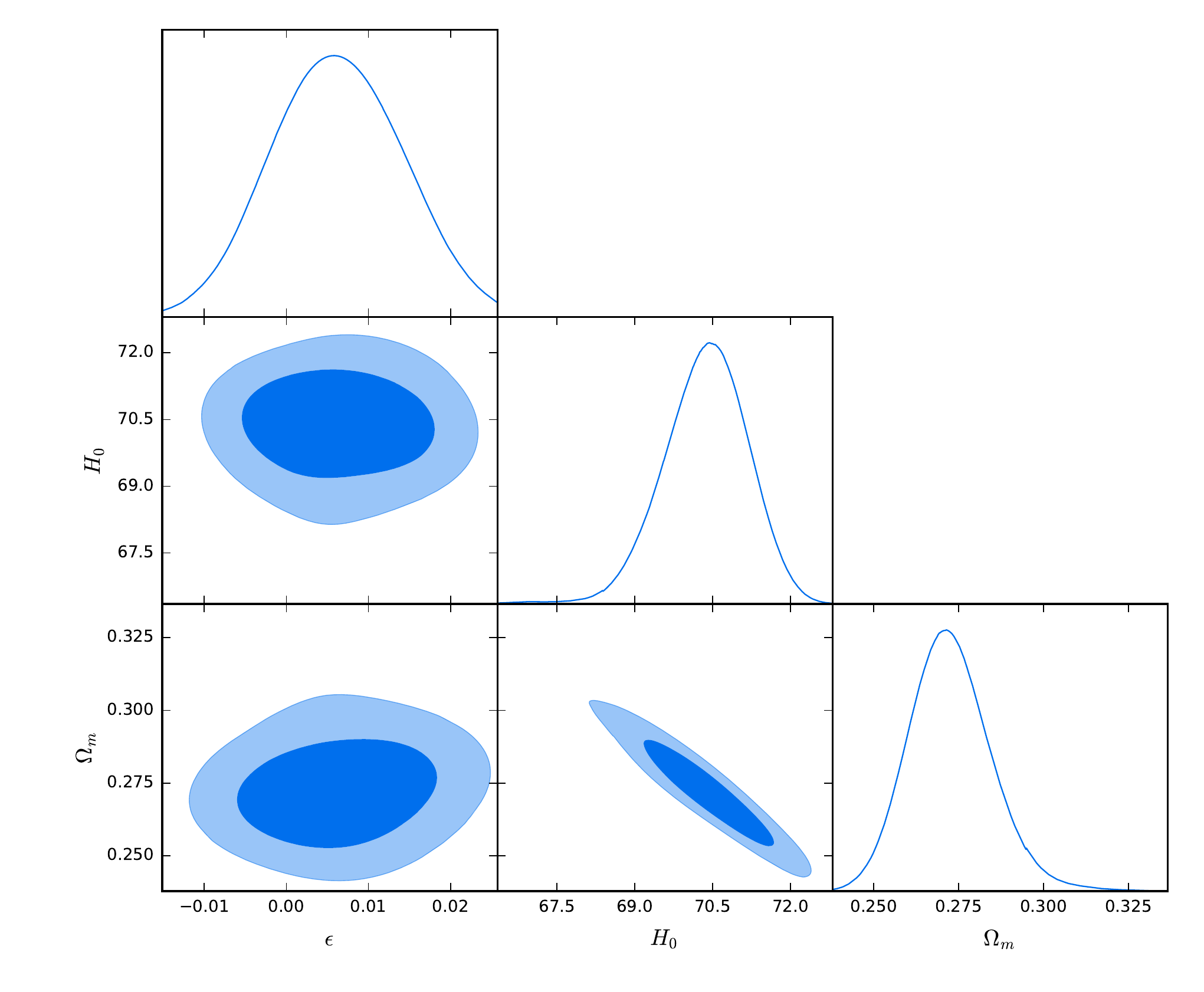}
\caption{For the first ADD sample, we present the 1-dimensional marginalized posterior distributions and 2-dimensional contours for the parameters of the $\Lambda$CDM, non-flat $\Lambda$CDM, $\omega$CDM and DV models, respectively.}\label{f1}
\end{figure}

\begin{figure}
\centering
\includegraphics[scale=0.4]{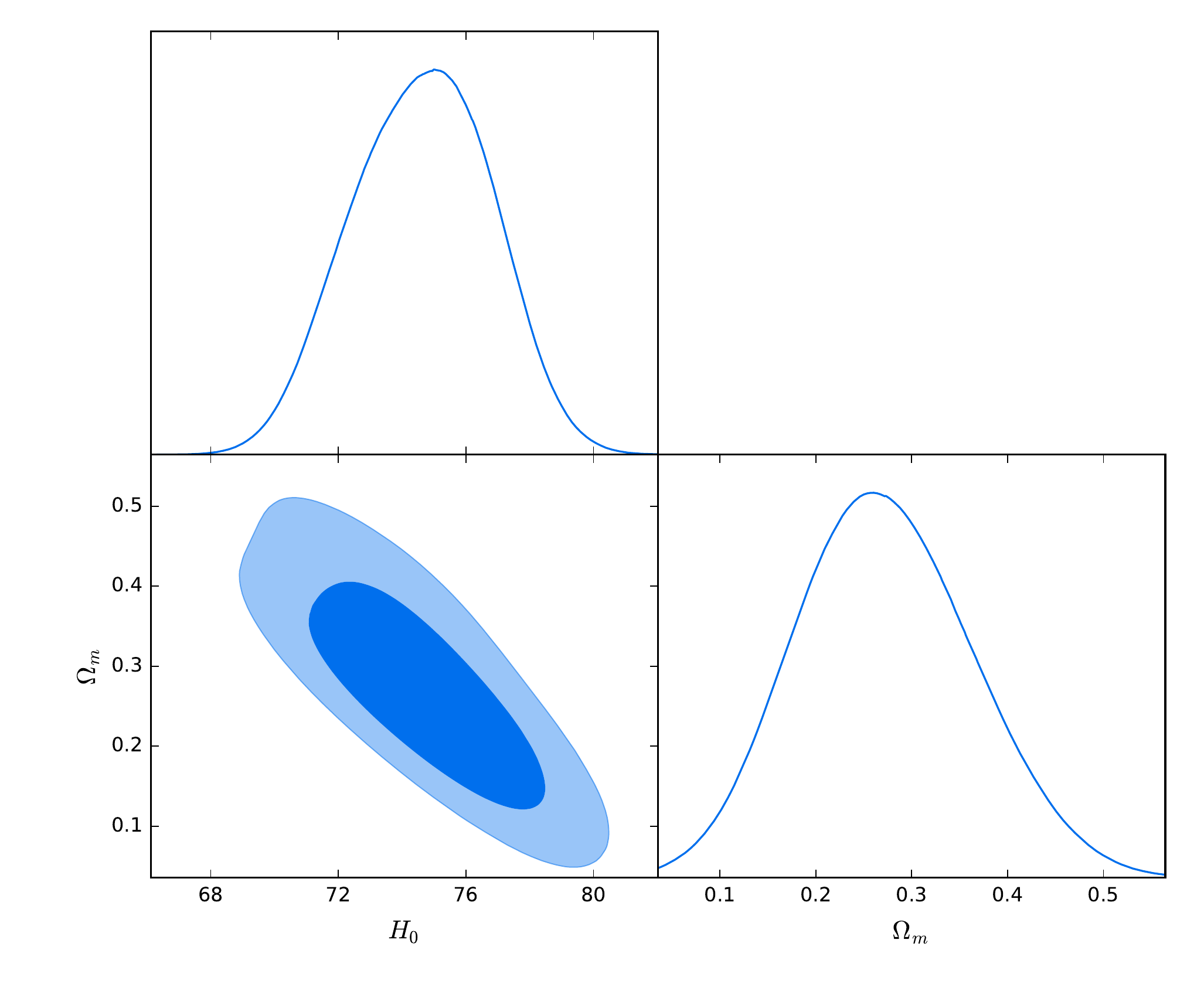}
\includegraphics[scale=0.4]{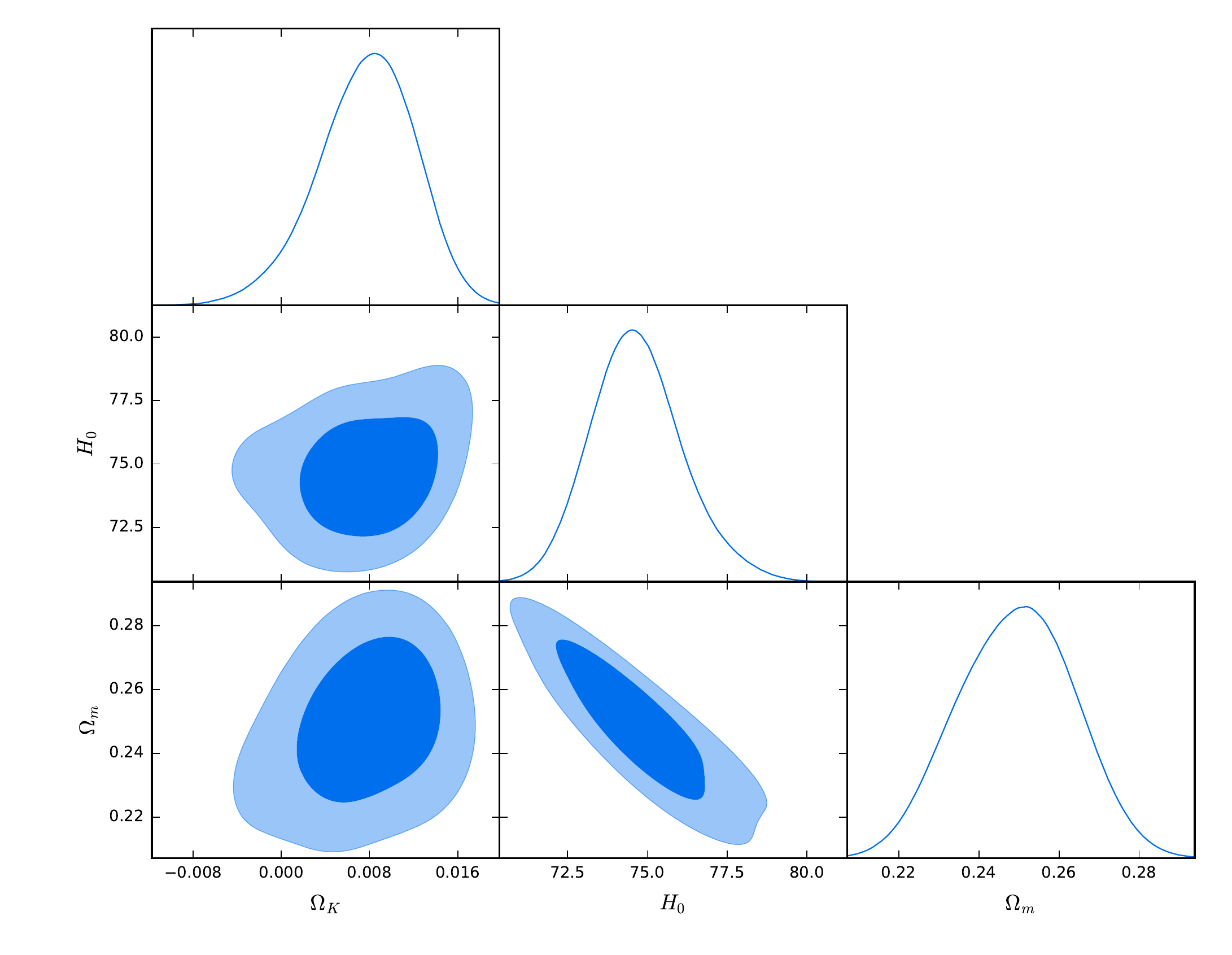}
\includegraphics[scale=0.4]{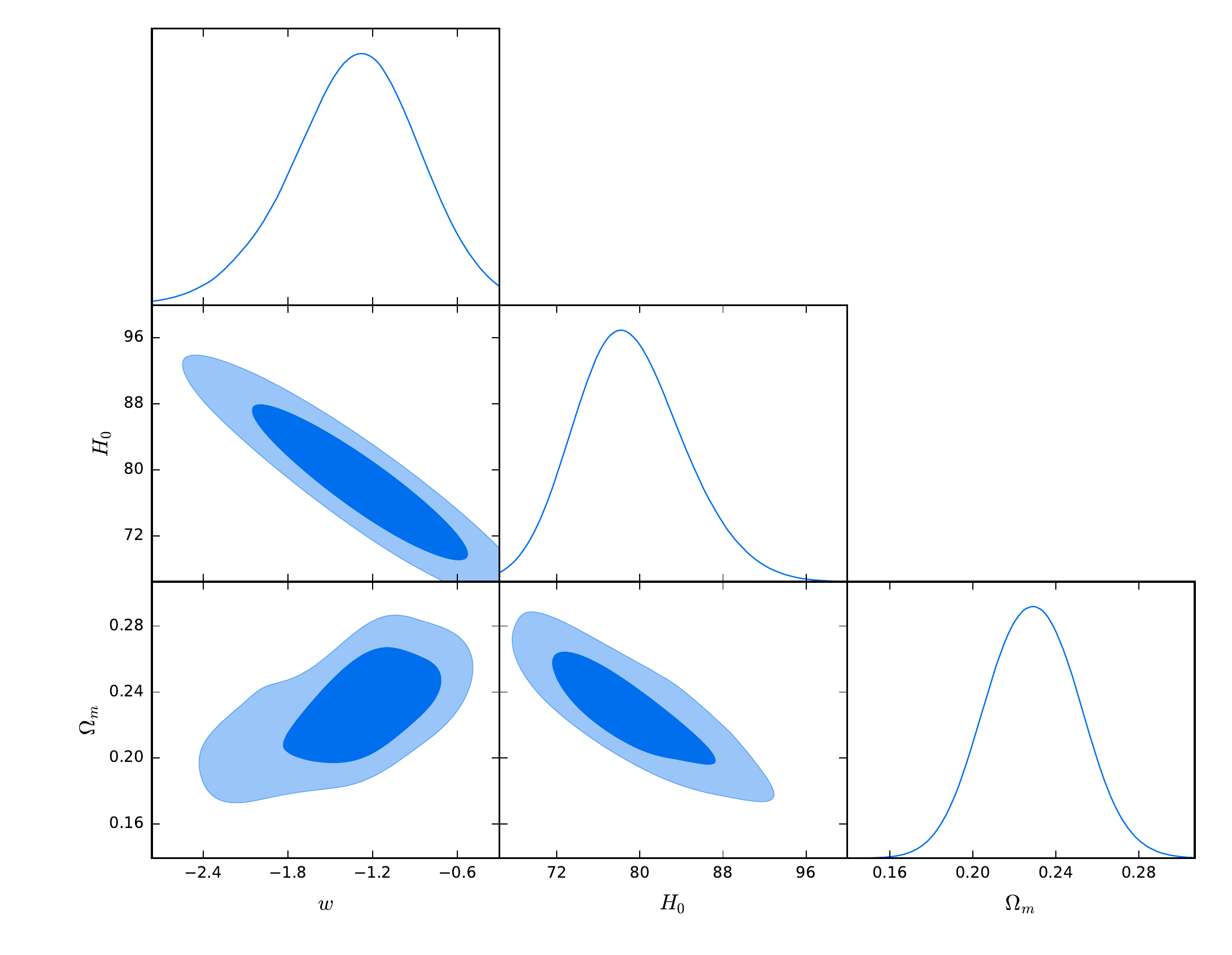}
\includegraphics[scale=0.4]{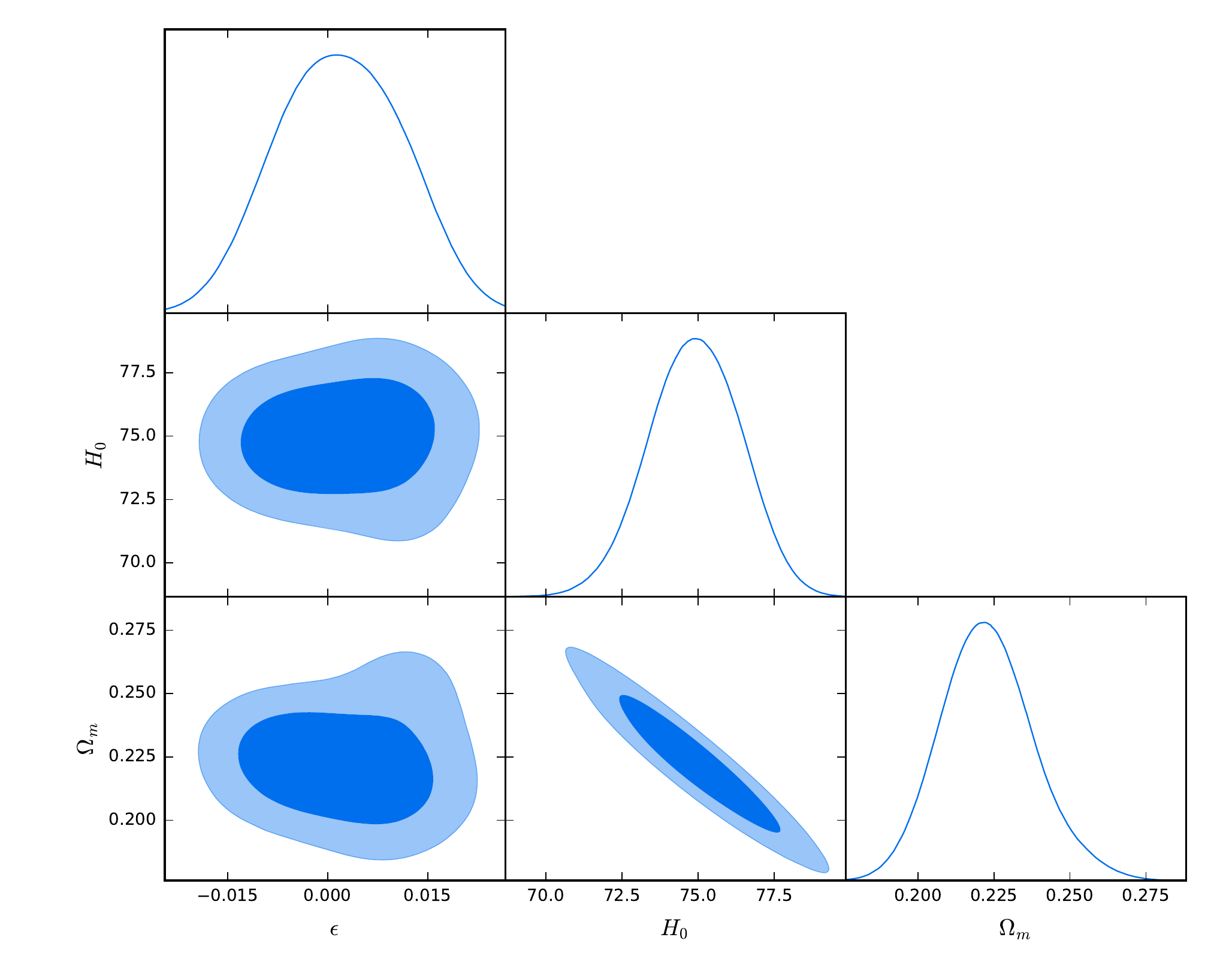}
\caption{For the second ADD sample, we present the 1-dimensional marginalized posterior distributions and 2-dimensional contours for the parameters of the $\Lambda$CDM, non-flat $\Lambda$CDM, $\omega$CDM and DV models, respectively.}\label{f2}
\end{figure}

\begin{table}[h!]
\renewcommand\arraystretch{1.3}
\begin{tabular}{ll}

\hline
\hline
Parameters           &Priors                                         \\
\hline
$H_0$                &$[20, 100]$                                   \\
$\Omega_{m}$        &$[0, 0.99]$                                 \\
$\Omega_{k}$        &$[-0.05, 0.05]$                                             \\
$\omega$             &$[-3, 1]$                                      \\
$\epsilon$           &$[-0.03, 0.03]$                                                 \\
\hline
\hline
\end{tabular}
\caption{The priors of different model parameters used in the Bayesian analysis.}
\label{t1}
\end{table}

\begin{table}[h!]
\renewcommand\arraystretch{1.3}
\begin{tabular}{ccccccc}

\hline
\hline
Parameter            &$\Lambda$CDM            &non-flat $\Lambda$CDM        &$\omega$CDM            &DV                               \\
\hline
$H_0$                &$70.1\pm 0.8$ &$69.7\pm 1.0 $      &$70.9^{+1.1}_{-1.2}$ &$70.3^{+0.8}_{-0.7}$                           \\
$\Omega_{m}$        &$0.283\pm 0.008$ &$0.320^{+0.064}_{-0.077}$  &$0.283\pm 0.011$ &$0.272^{+0.011}_{-0.013}$                            \\
$\Omega_{k}$             &---        &$0.0151^{+0.0174}_{-0.0202}$                                   &---               &---                         \\                                                  $\omega$             &---                        &---                      &$-1.081\pm 0.048$           &---                                       \\
$\epsilon$           &---                        &---                      &---            &$0.0100^{+0.0726}_{-0.0754}$                                       \\
\hline
$\chi^2_{min}$       &1.2                          &1.4                         &1.9             &1.5                   \\
n                    &1                       &2                  &2                 &2 \\
$\Delta$AIC$_{c}$                  &0    &0.38     &0.88  &0.48        \\
$\Delta$BIC                        &0    &3.42     &3.92  &3.52              \\
\hline
\hline
\end{tabular}
\caption{The constraining results for four cosmological models using the first ADD sample: the $1\sigma$ (68\%) confidence ranges of free parameters and values of two information criterions.}
\label{t2}
\end{table}

\begin{table}[h!]
\renewcommand\arraystretch{1.3}
\begin{tabular}{ccccccc}

\hline
\hline
Parameter            &$\Lambda$CDM            &non-flat $\Lambda$CDM        &$\omega$CDM            &DV                               \\
\hline
$H_0$                &$74.6\pm 2.1$ &$74.7^{+1.3}_{-1.5}$      &$79.0^{+4.7}_{-5.6}$ &$74.9\pm 1.5 $                           \\
$\Omega_{m}$        &$0.272^{+0.086}_{-0.099}$ &$0.249\pm 0.014$  &$0.229\pm 0.021$ &$0.223^{+0.014}_{-0.016}$                            \\
$\Omega_{k}$             &---        &$0.0078^{+0.0047}_{-0.0040}$                                   &---               &---                         \\                                                  $\omega$             &---                        &---                      &$-1.342^{+0.491}_{-0.418} $           &---                                       \\
$\epsilon$           &---                        &---                      &---            &$0.0019\pm 0.0089$                                       \\
\hline
$\chi^2_{min}$       &104.5                           &104.8                        &105.6             &105.2                   \\
n                    &1                       &2                  &2                 &2 \\
$\Delta$AIC$_{c}$                  &0    &0.48    &1.28  &0.88        \\
$\Delta$BIC                        &0    &3.51     &4.31  &3.91              \\
\hline
\hline
\end{tabular}
\caption{The constraining results for four cosmological models using the second ADD sample: the $1\sigma$ (68\%) confidence ranges of free parameters and values of two information criterions.}
\label{t3}
\end{table}
\begin{figure}
\centering
\includegraphics[scale=0.4]{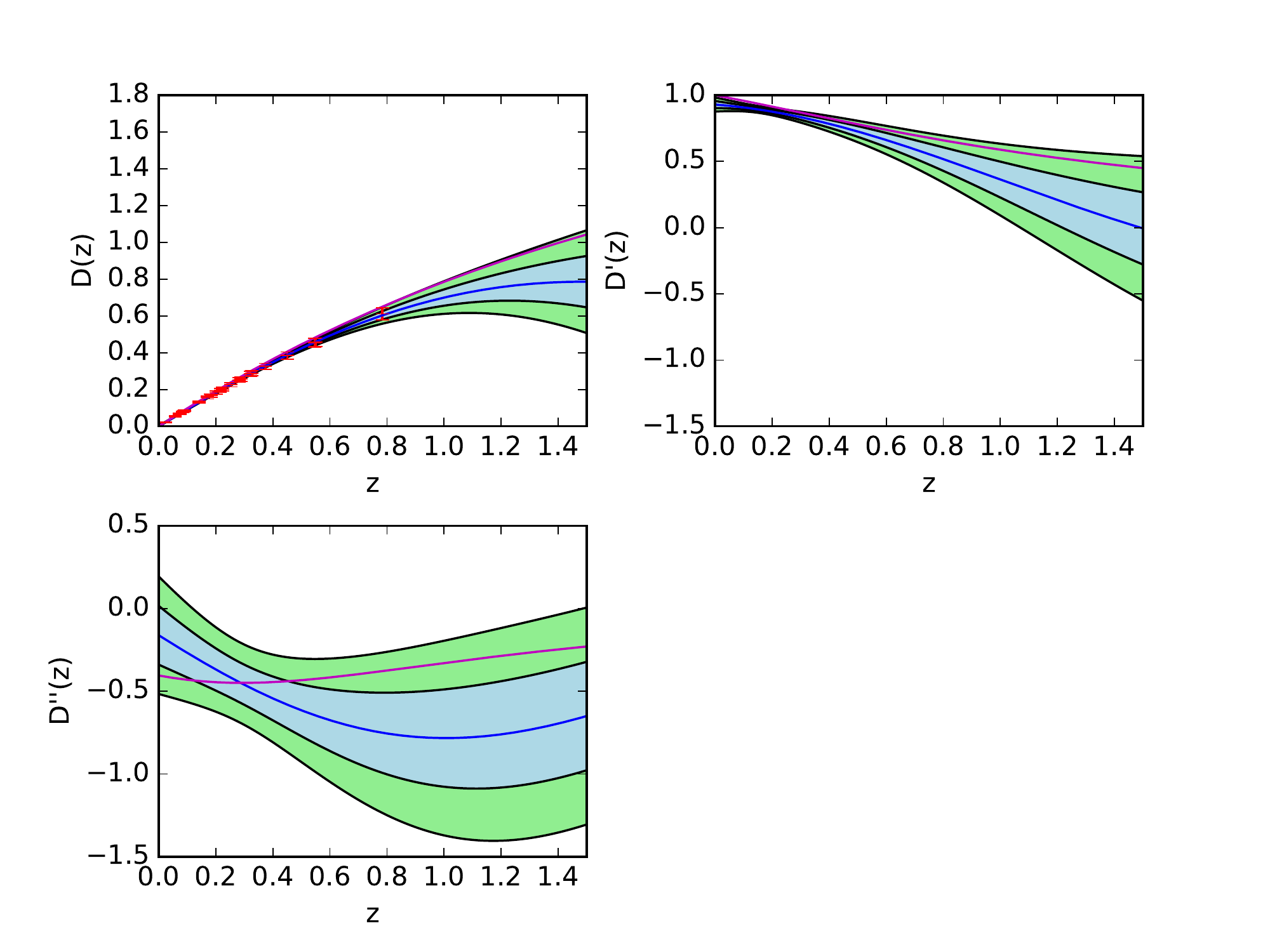}
\includegraphics[scale=0.4]{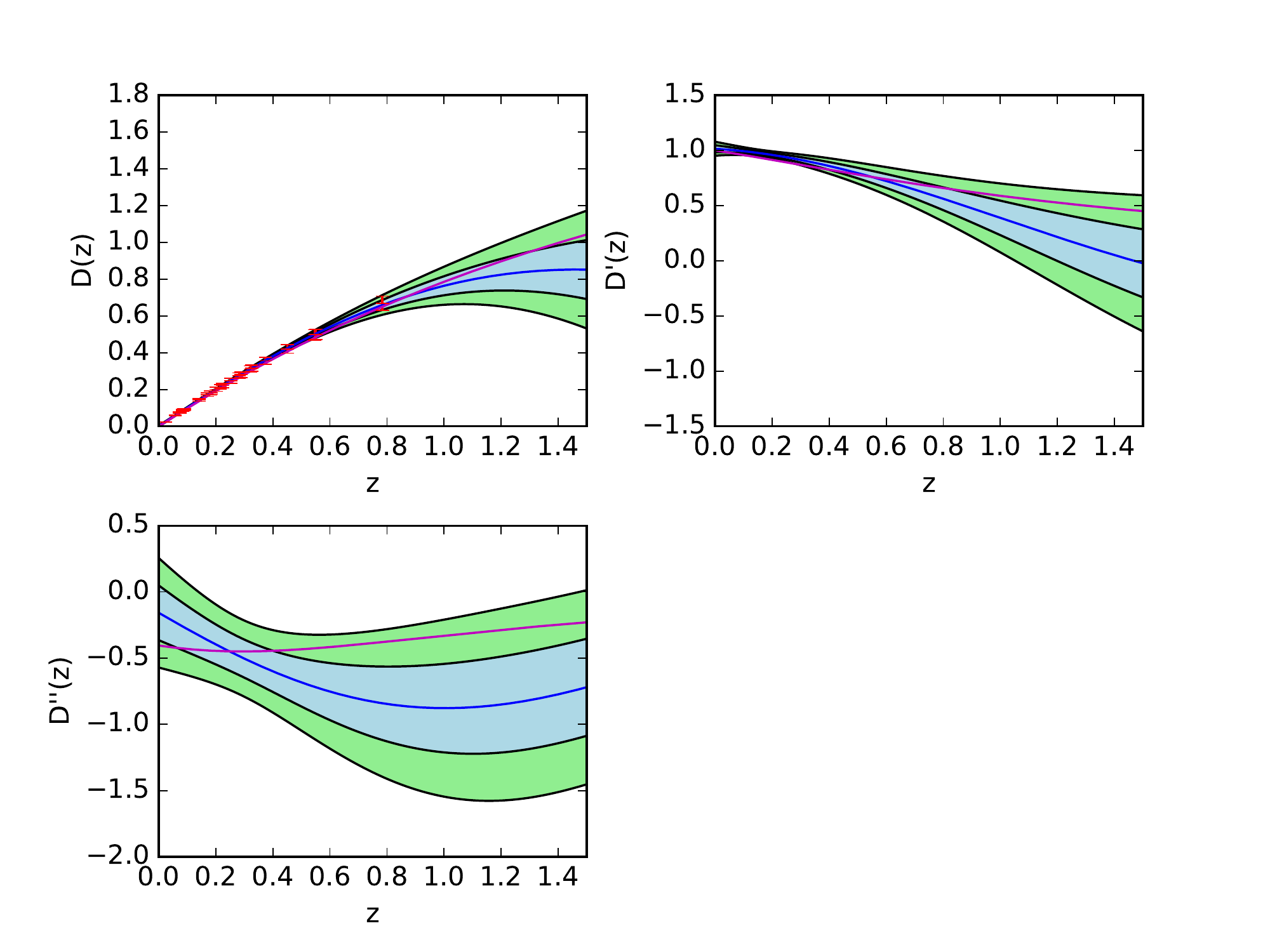}
\caption{The GP reconstructions of $D(z), D'(z)$ and $D''(z)$ using the first ADD sample. The left and right panels correspond to the cases of $H_0=66.93\pm0.62$ and $73.24\pm1.74$ km s$^{-1}$ Mpc$^{-1}$, respectively. The points with red error bars are 25 ADD measurements. The shaded regions are reconstructions with $68\%$ and $95\%$ CL. The blue and magenta lines denote the underlying true model and the $\Lambda$CDM model, respectively. }\label{f3}
\end{figure}
\begin{figure}
\centering
\includegraphics[scale=0.28]{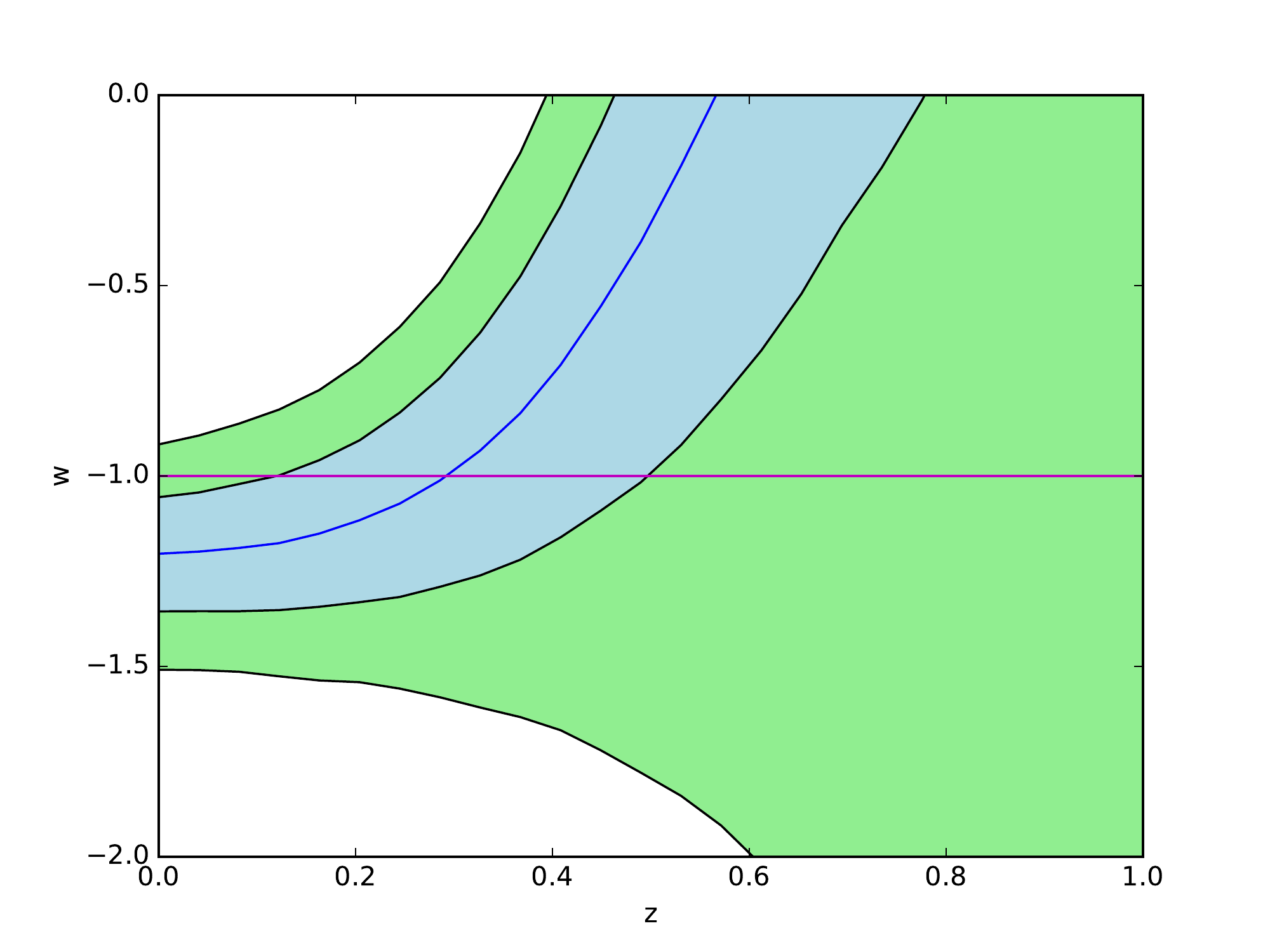}
\includegraphics[scale=0.28]{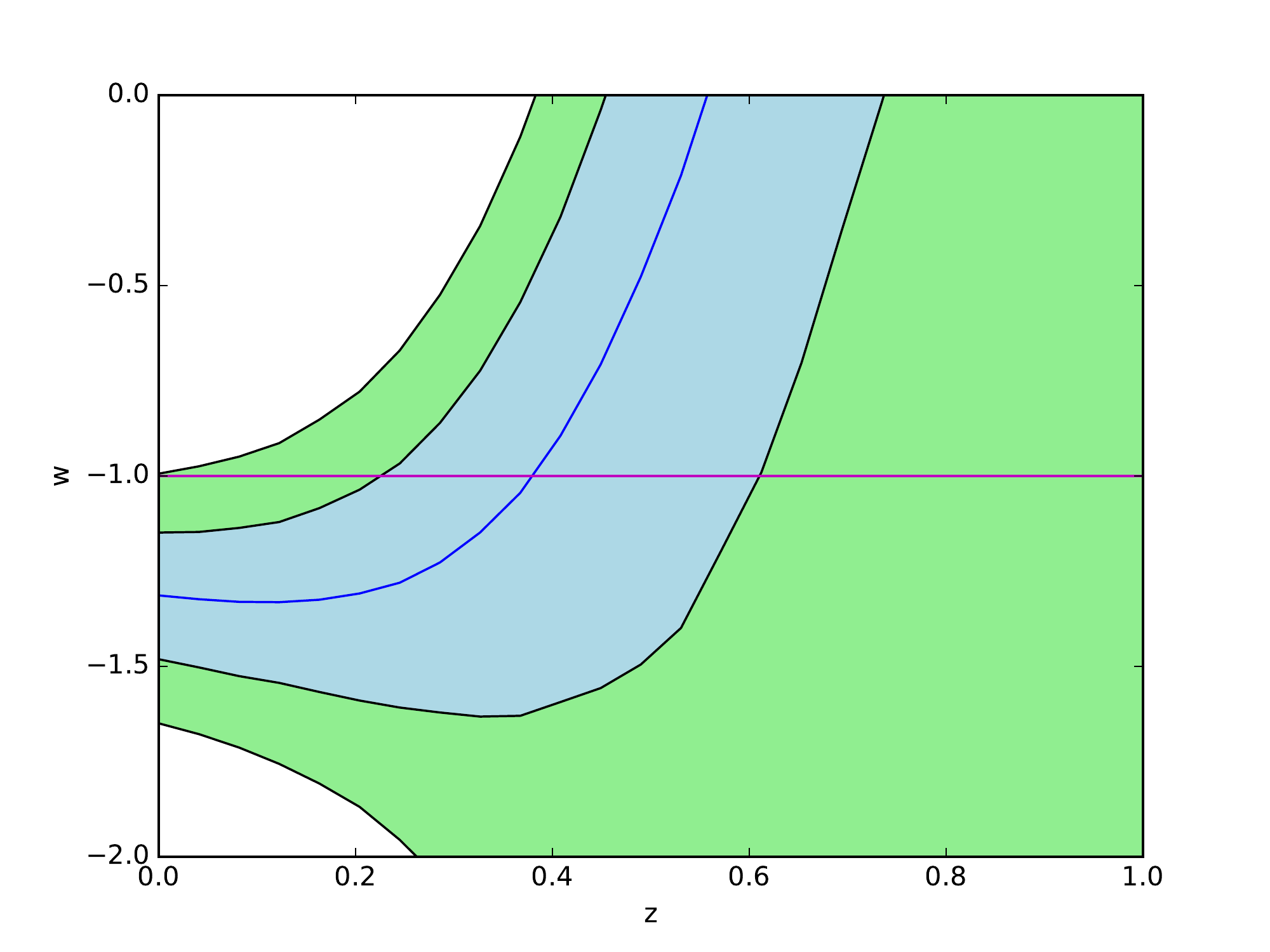}
\includegraphics[scale=0.28]{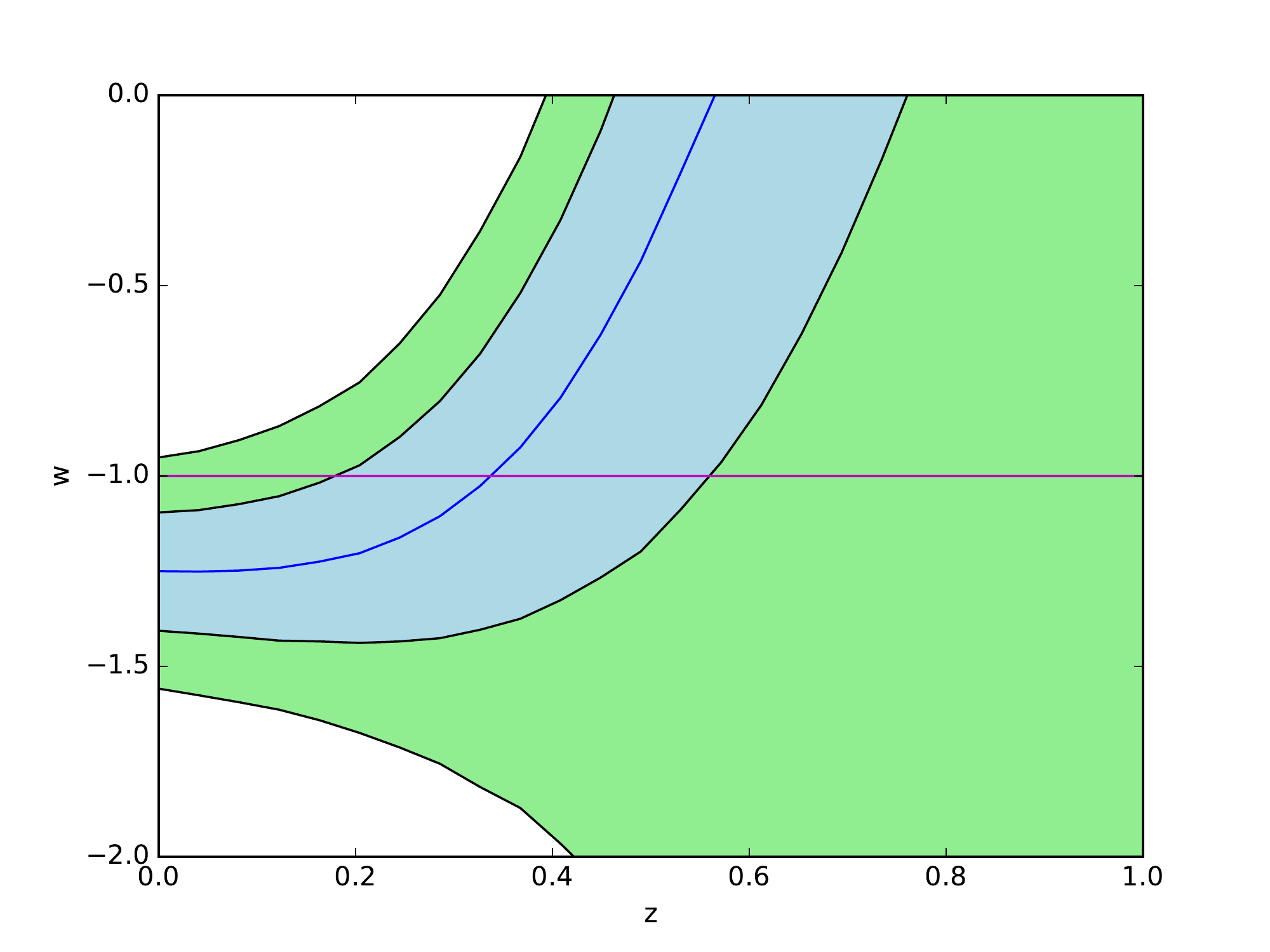}
\caption{The GP reconstructions of the EoS of DE $\omega(z)$ using the first ADD sample. From left to right, different panels correspond to the cases of $H_0=66.93\pm0.62$, $73.24\pm1.74$ and $70.1\pm0.8$ ($\Lambda$CDM) km s$^{-1}$ Mpc$^{-1}$, respectively.}\label{f4}
\end{figure}
\begin{figure}
\centering
\includegraphics[scale=0.4]{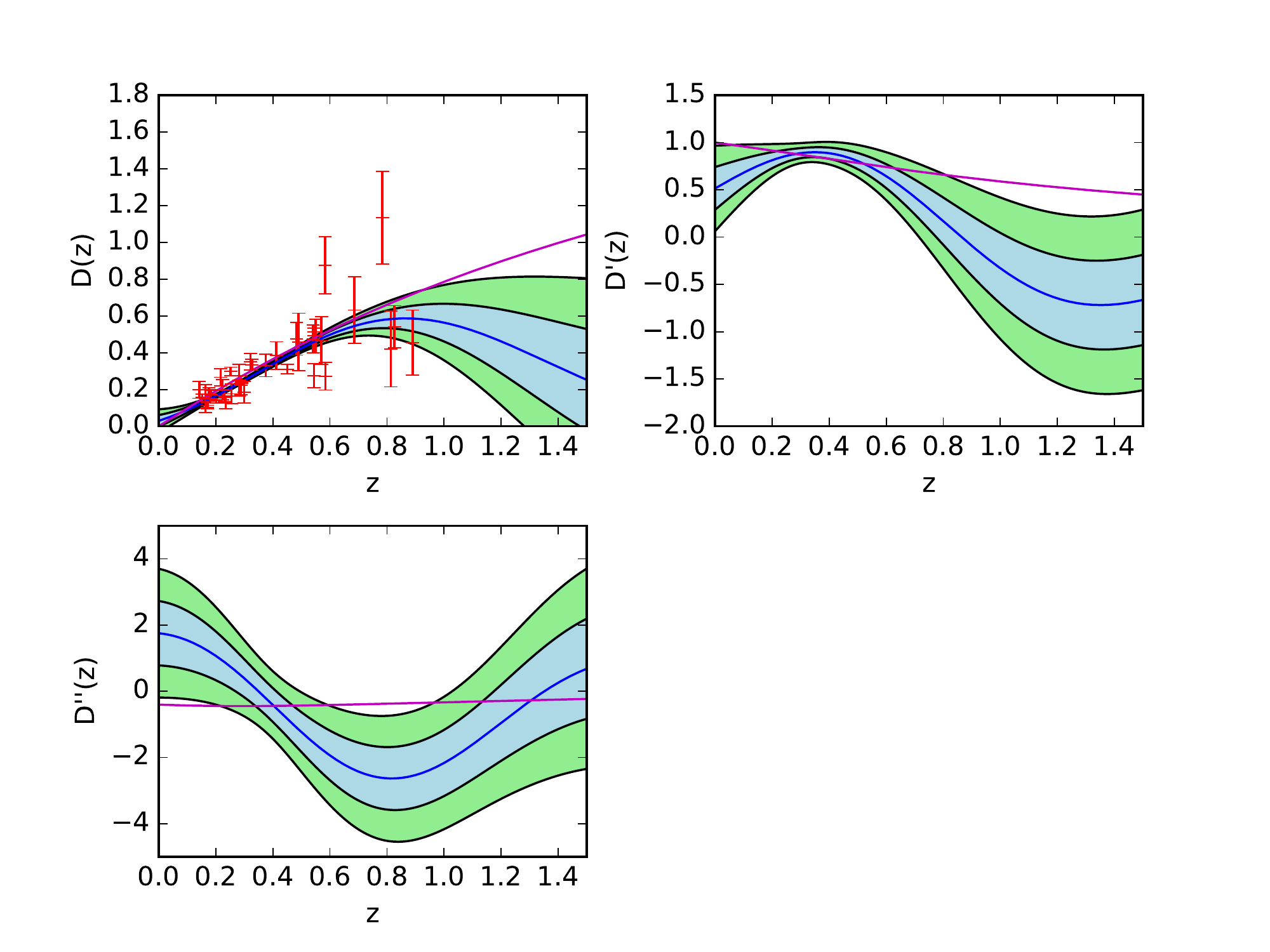}
\includegraphics[scale=0.4]{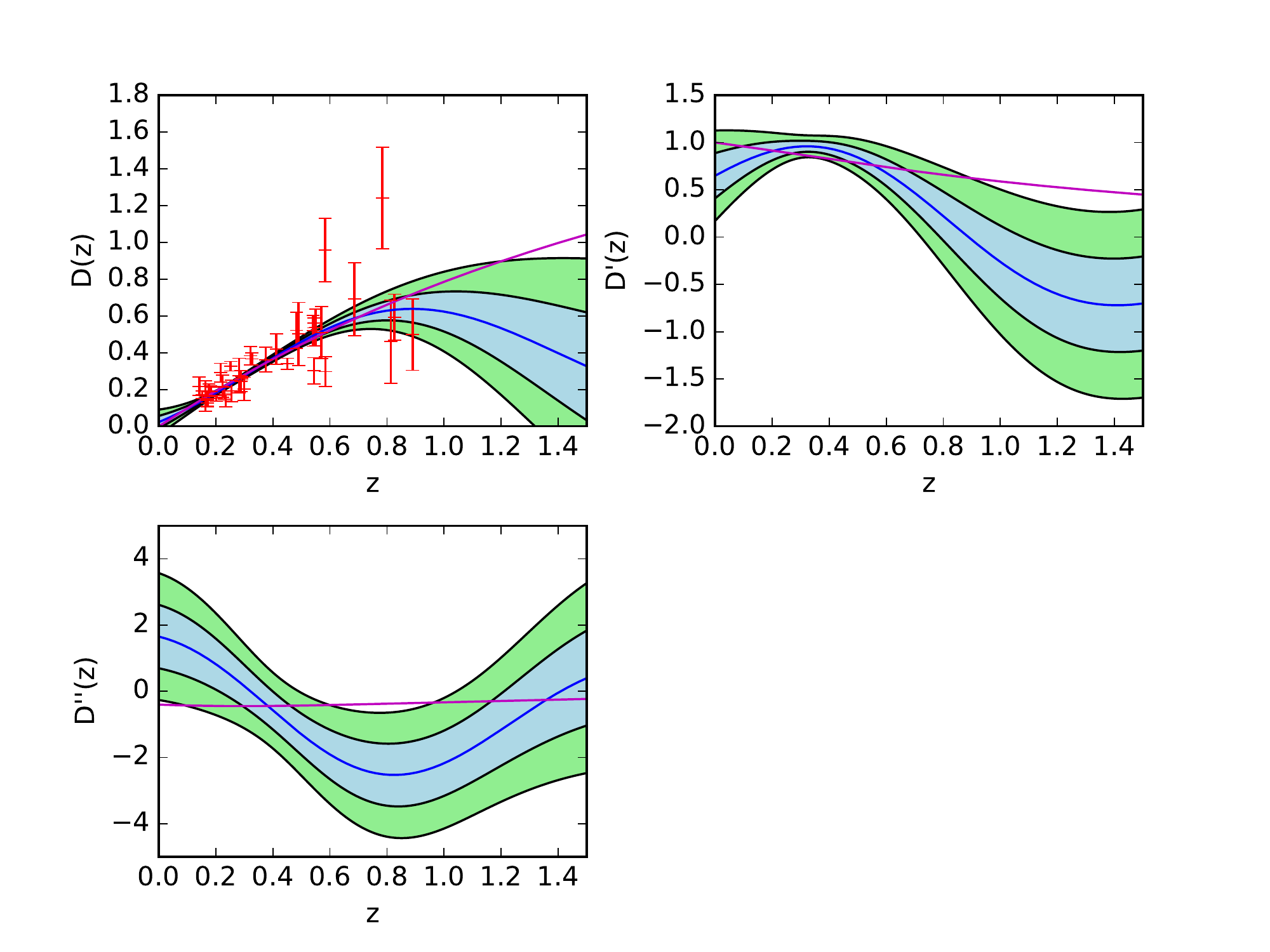}
\caption{The GP reconstructions of $D(z), D'(z)$ and $D''(z)$ using the second ADD sample. The left and right panels correspond to the cases of $H_0=66.93\pm0.62$ and $73.24\pm1.74$ km s$^{-1}$ Mpc$^{-1}$, respectively. The points with red error bars are 38 ADD measurements.}\label{f5}
\end{figure}
\begin{figure}
\centering
\includegraphics[scale=0.3]{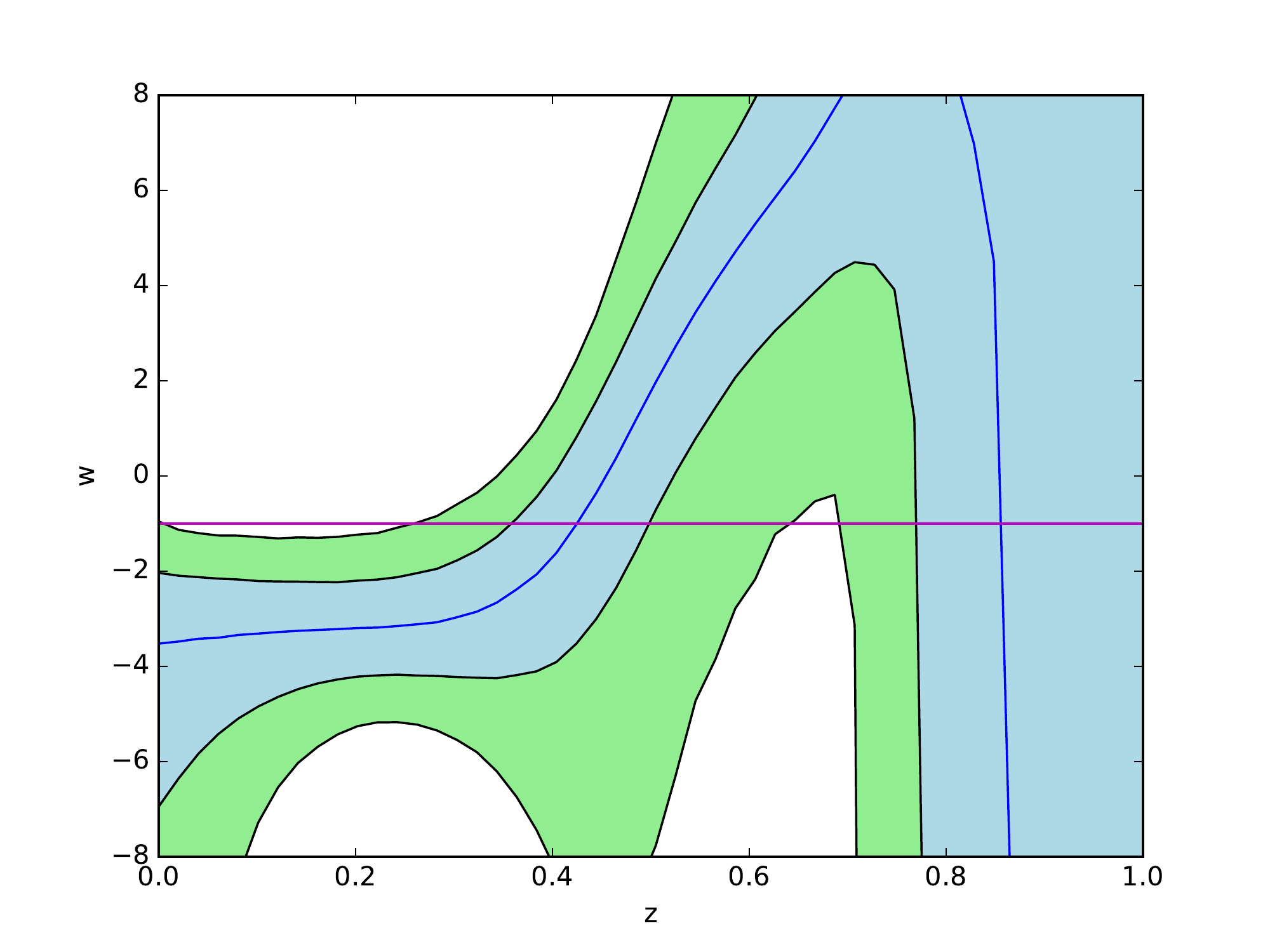}
\includegraphics[scale=0.3]{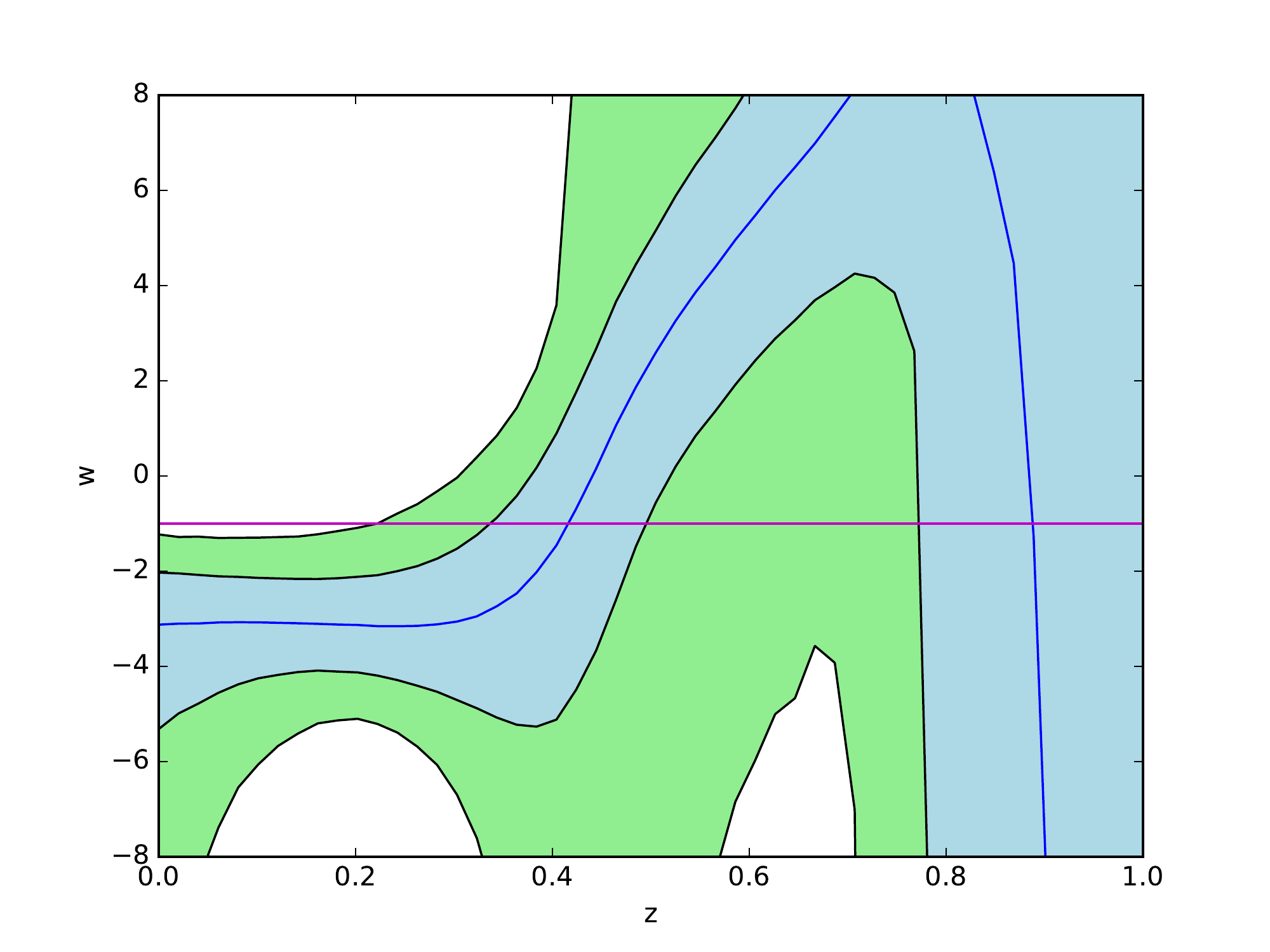}
\includegraphics[scale=0.3]{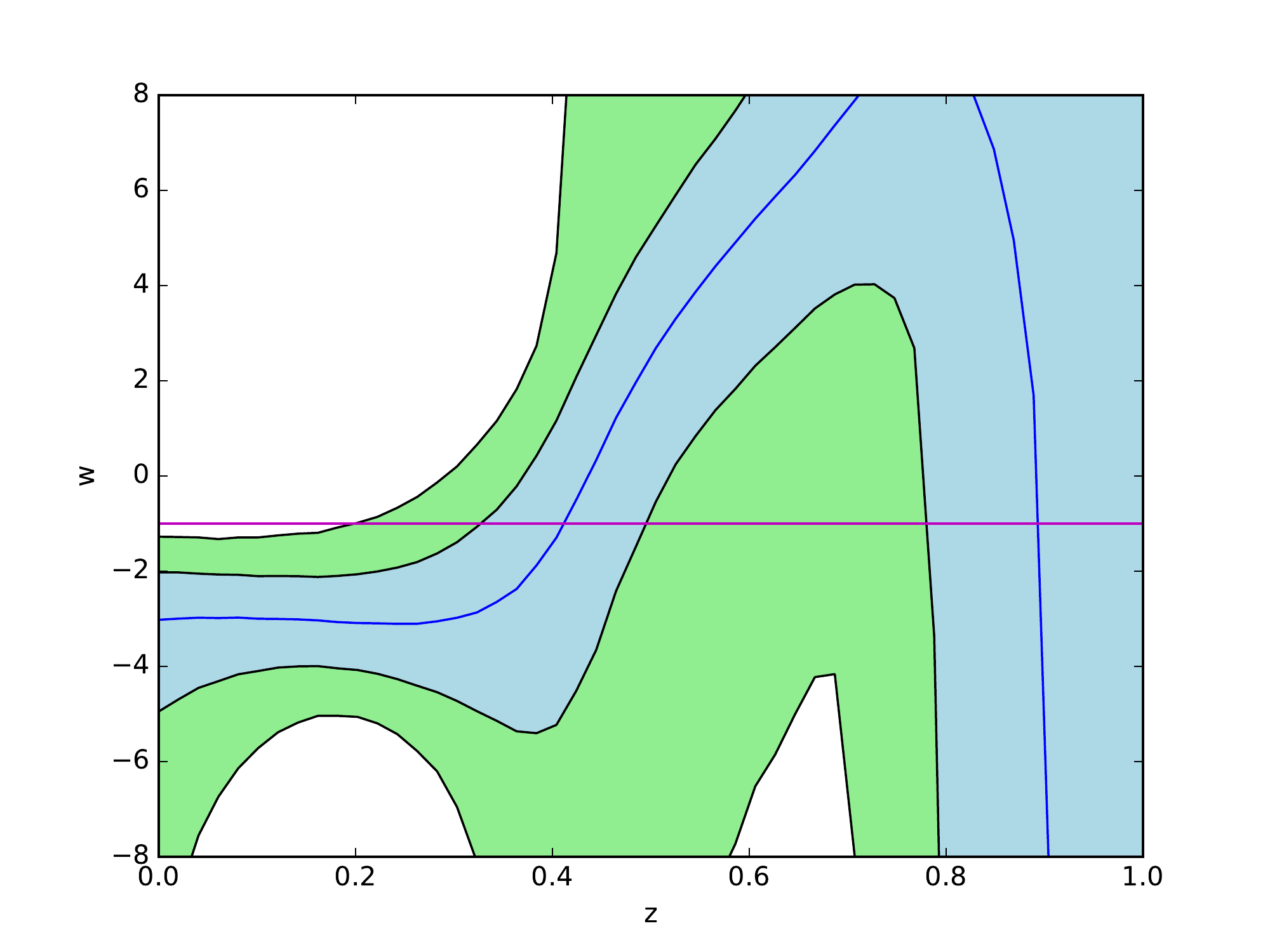}
\includegraphics[scale=0.3]{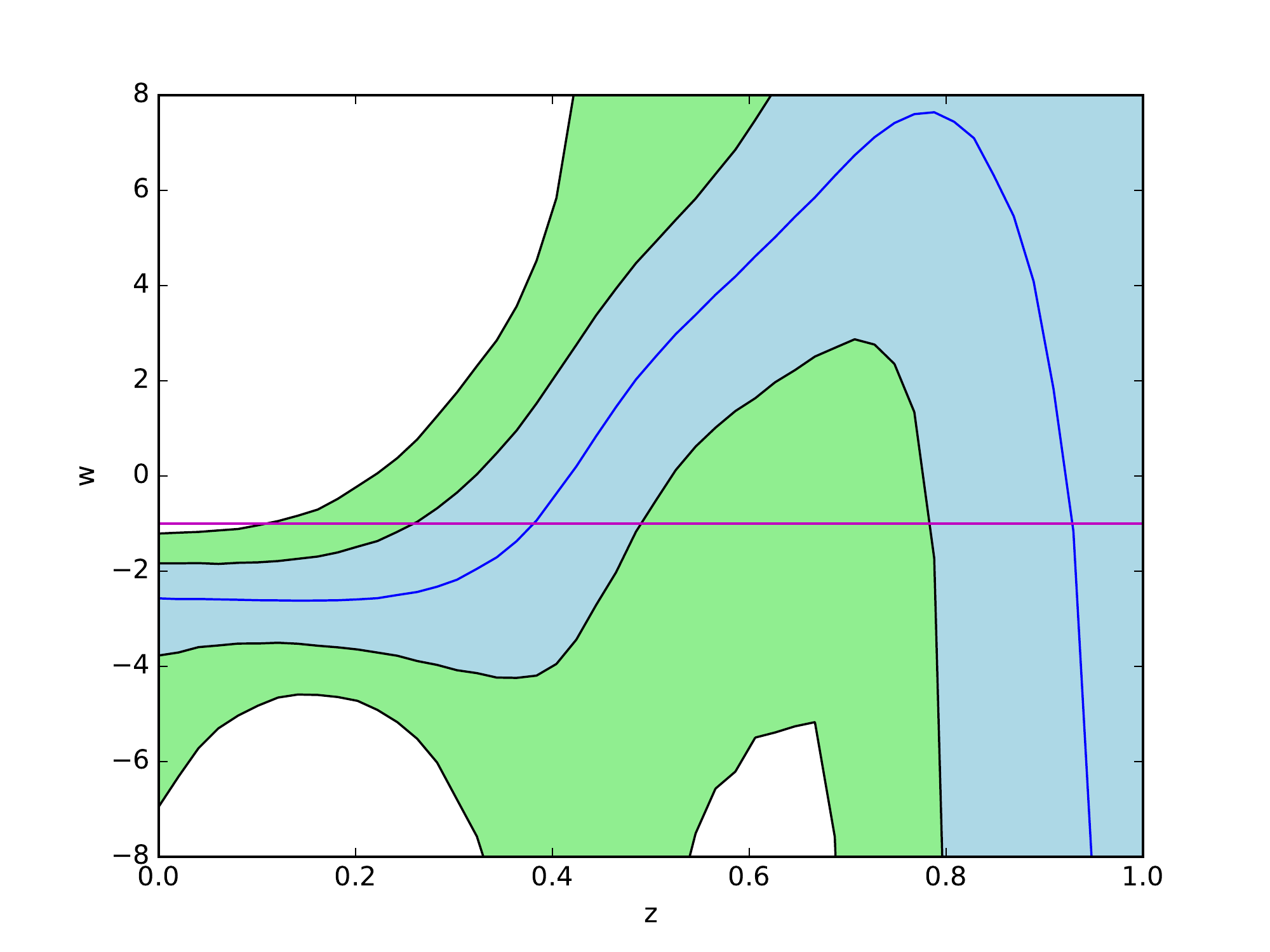}
\caption{The GP reconstructions of the EoS of DE $\omega(z)$ using the second ADD sample. From left to right, different panels correspond to the cases of $H_0=66.93\pm0.62$, $73.24\pm1.74$, $74.6\pm2.1$ ($\Lambda$CDM) and $79.0^{+4.7}_{-5.6}$ ($\omega$CDM) km s$^{-1}$ Mpc$^{-1}$ in turn, respectively.}\label{f6}
\end{figure}
\begin{figure}
\centering
\includegraphics[scale=0.3]{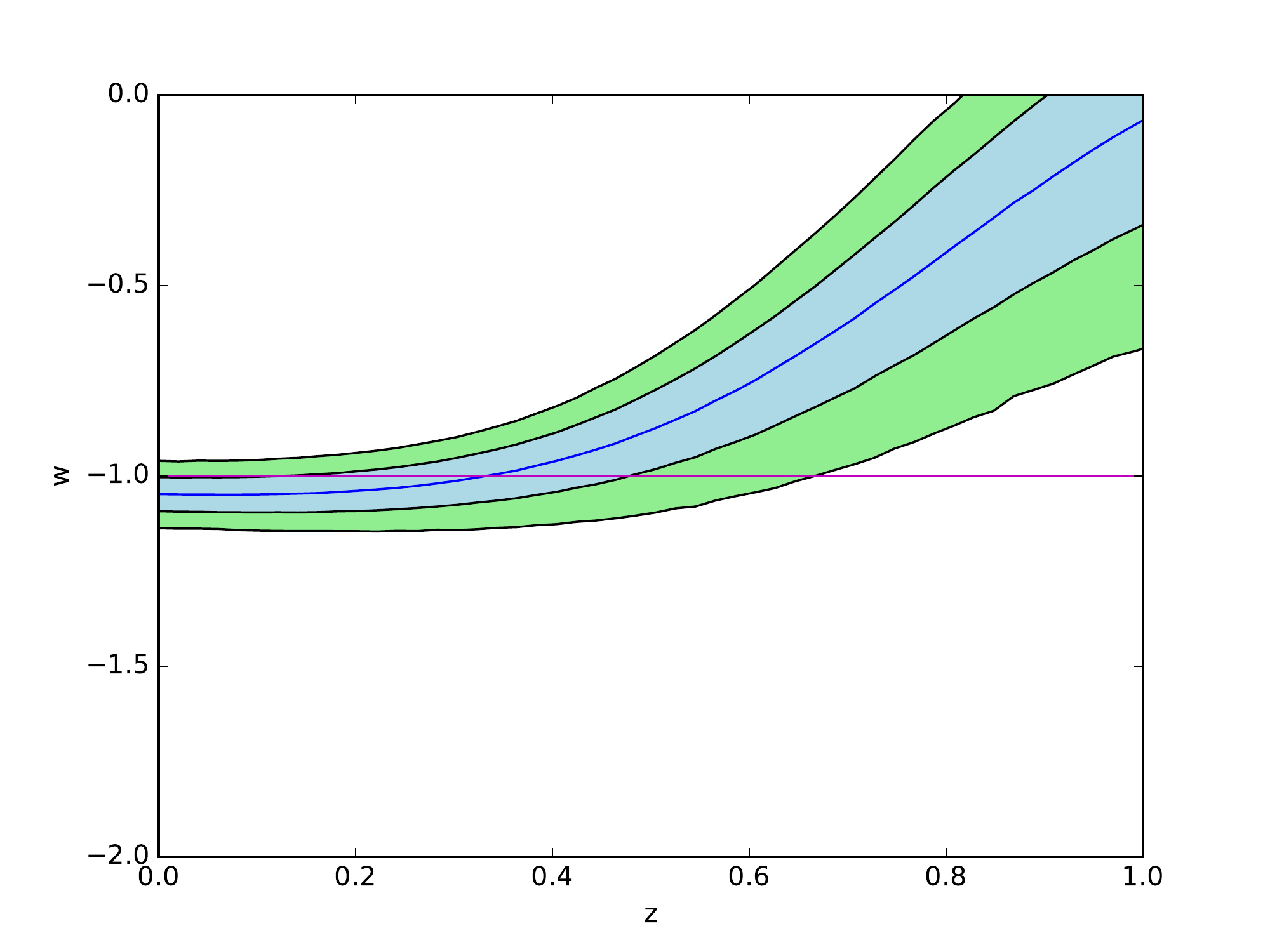}
\includegraphics[scale=0.3]{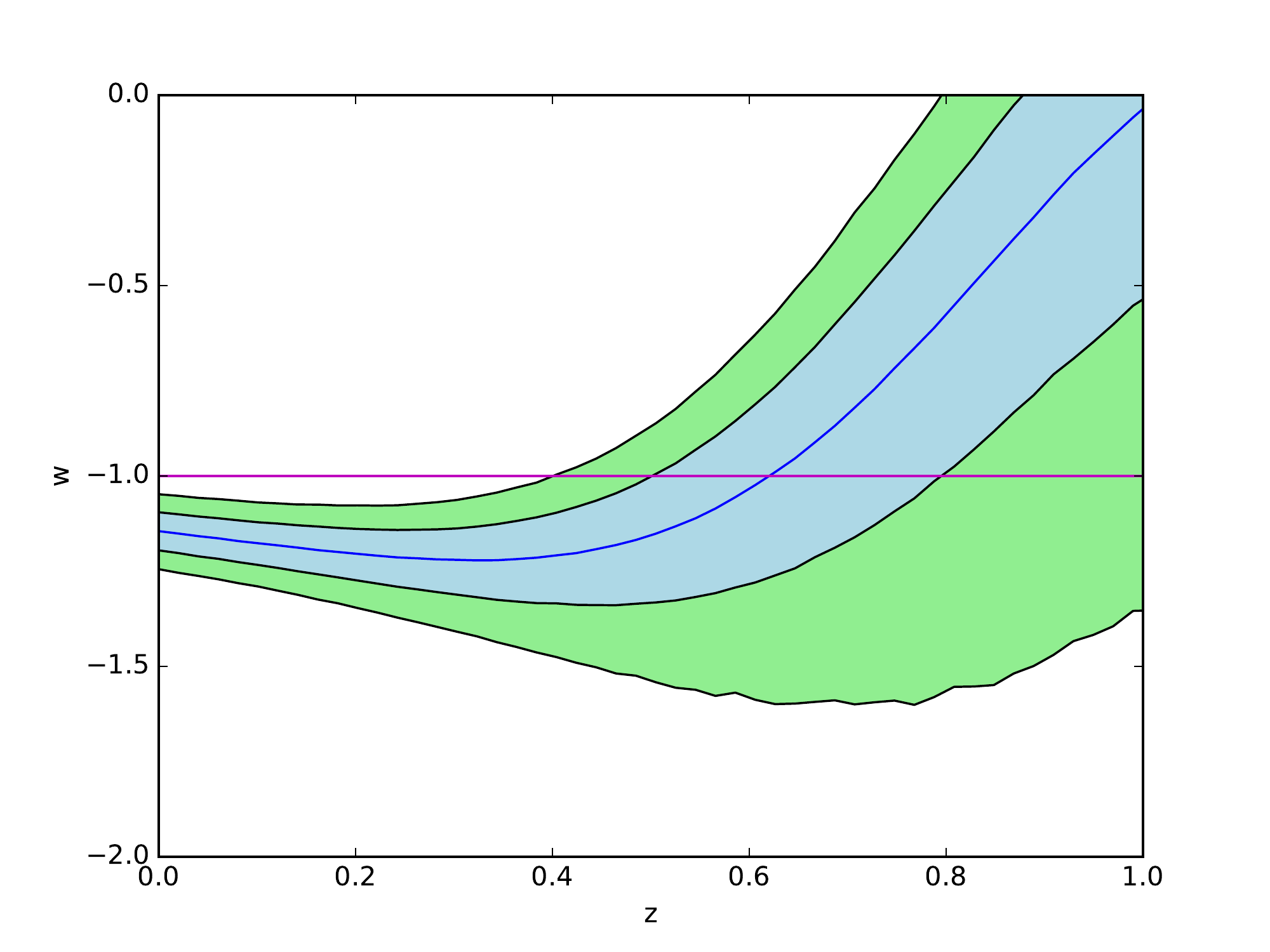}
\includegraphics[scale=0.3]{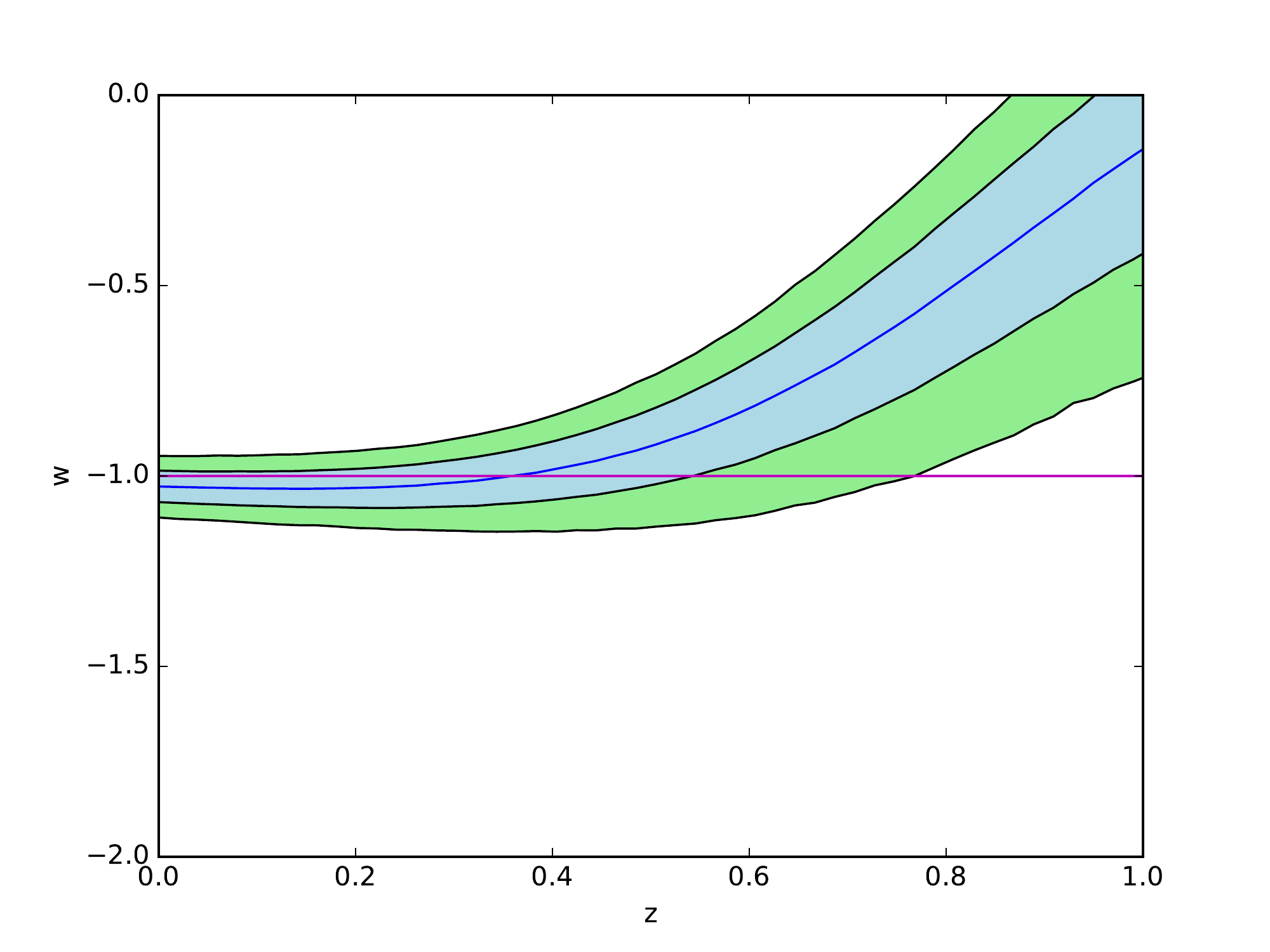}
\includegraphics[scale=0.3]{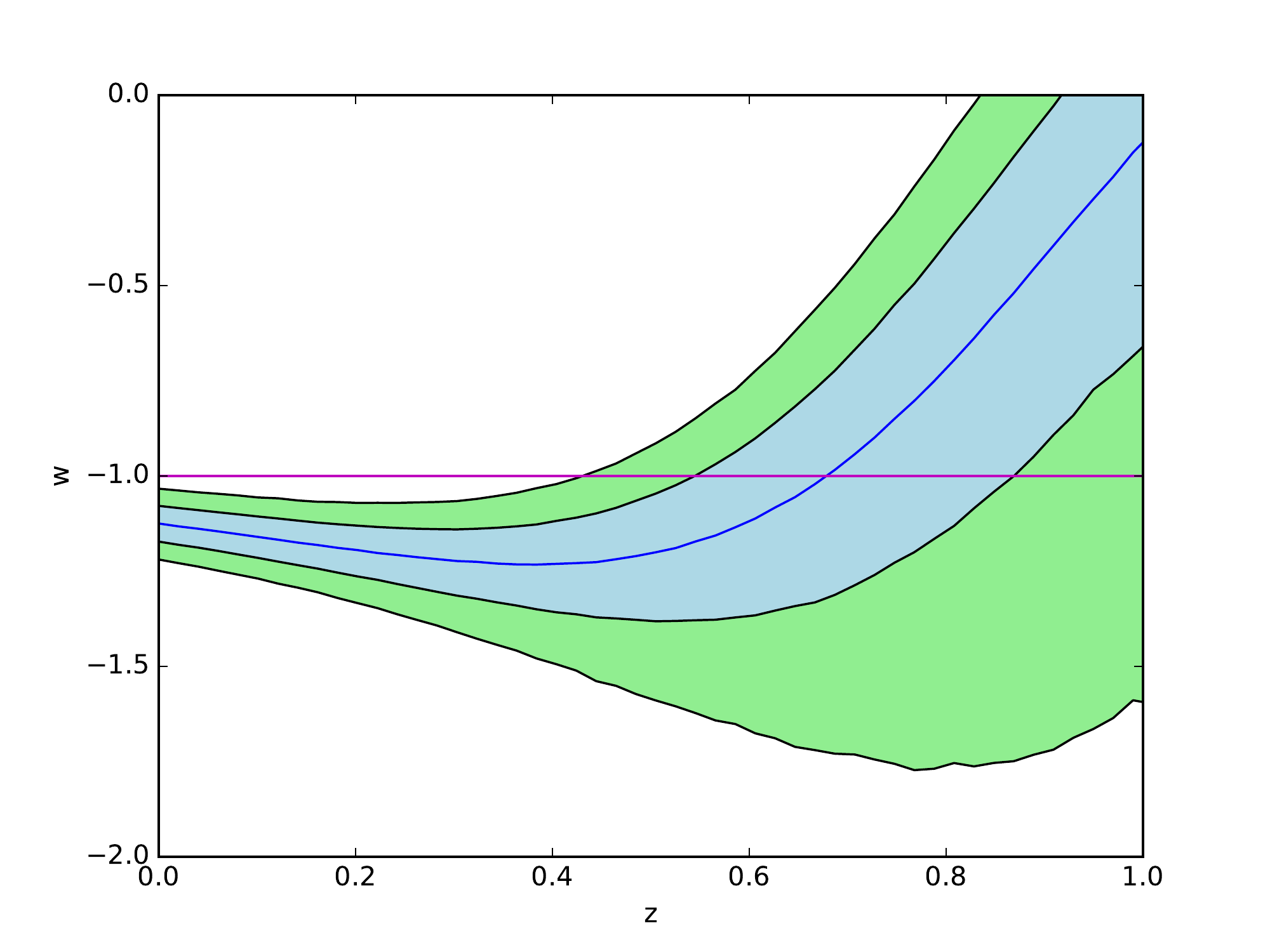}
\caption{The GP reconstructions of the EoS of DE $\omega(z)$ using a combination of ADD, SNe Ia, cosmic chronometers, Planck-2015 shift parameter and HII galaxy measurements. From left to right, the upper two panels correspond to the cases of $H_0=66.93\pm0.62$ and $73.24\pm1.74$ km s$^{-1}$ Mpc$^{-1}$ for the first ADD sample, respectively. Similarly, the lower two panels correspond to the cases of $H_0=66.93\pm0.62$ and $73.24\pm1.74$ km s$^{-1}$ Mpc$^{-1}$ for the second ADD sample, respectively.}\label{f7}
\end{figure}

To implement the Bayesian analysis, we use the Markov Chain Monte Carlo (MCMC) method to minimize the values of $\chi^2$. In particular, we modify carefully the public package CosmoMC \cite{a1} and  Boltzmann code CAMB \cite{a2}. We also list the priors of free parameters of four cosmological models in Table. \ref{t1}. Furthermore, in order to make a comparison between different models, we employ two selection criterions, i.e., Akaike information criterion (AIC) \cite{20} and Bayesian information criterion (BIC) \cite{21}, to assess these four cosmological models. These two criterions favor the models which have fewer parameters when giving the same fit, and have been widely used in the literature \cite{22,23,24,25}.

The AIC is usually defined as $\mathrm{AIC}=\chi^2_{min}+2n$. In this situation, since these two ADD sample sizes are smaller than the usual SNe Ia sample size, we utilize a modified version of the AIC, $\mathrm{AIC}_c=\mathrm{AIC}+2n(n-1)/(N-n-1)$, which is very efficient for the ratio $N/n\lesssim40$, where $n$ denotes the number of free parameters (for details, see \cite{22}). The BIC is usually defined as $\mathrm{BIC}=\chi^2_{min}+n\ln N$. It is noteworthy that the relative values of these two criterions between different models are more useful than the absolute values to distinguish which model is preferred by the observed data. Therefore, we define the relative variations $\Delta \mathrm{AIC_{c}}$ and $\Delta \mathrm{BIC}$ of the two criterions between models as follows
\begin{equation}
\Delta \mathrm{AIC}_c=\mathrm{AIC}_c^{model}-\mathrm{AIC}_c^{\Lambda CDM}, \label{9}
\end{equation}
\begin{equation}
\Delta \mathrm{BIC}=\mathrm{BIC}^{model}-\mathrm{BIC}^{\Lambda CDM}, \label{10}
\end{equation}
where we have chosen the $\Lambda$CDM model as the reference model. In general, the BIC is more stringent than the AIC for the models with extra parameters, when the data sample satisfies $\ln N >2$. More specifically, the differences $\Delta \mathrm{BIC}\sim2$ and $\Delta \mathrm{BIC}\sim6$ indicate, respectively, a positive and strong evidence against the reference model. Notice that here the IC results only reflect the preference of current data and it is possible that future high-quality ADD data can give different constraining results for these DE models.

For the first and second ADD samples, the best-fit values of these model parameters, the minimal values of the derived $\chi^2$ and the diagnostic results using the information criterions are listed in Tabs. \ref{t2}-\ref{t3}, respectively. The corresponding $1\sigma$ and $2\sigma$ confidence contours for different model parameter pairs are shown in Figs. \ref{f1}-\ref{f2}. We find that, for these two samples, the values of $H_0$ are more consistent with the R16's result than the P15's one, and the $\Lambda$CDM model is preferred (see Table. \ref{t2}-\ref{t3}, $\Delta \mathrm{BIC}>3$ for the left three models), although higher $H_0$ values in the second sample than that in the first one are obtained. It is worth noting that the higher $H_0$ values can be ascribed to different numbers and qualities of the observed ADD data, different observational techniques and systematics. Subsequently, we find that a flat universe is still preferred for the first sample. However, this is not the case for the second one, i.e., a positive cosmic curvature corresponding to negatively-curved 3-geometries is mildly supported by 38 ADD data points at the 1$\sigma$ CL. Moreover, we also obtain two common conclusions for both ADD samples: (i) the EoS of DE is consistent with $\omega=-1$ at the 1$\sigma$ CL; (ii) there is no indication of momentum transfer in the dark sector of the universe from constrained values of $\epsilon$, which are compatible with zero at the 1$\sigma$ CL.

In the following context, in the statistical framework of GP method, we try to search for the evidence of DDE using the ADD data.

\section{The GP}
A GP is a collection of random variables, any finite number of which have a joint Gaussian distribution \cite{26}. A GP can reconstruct a function from the observational data without assuming a special parametrization for the underlying function. At each reconstructed point $z$, the reconstructed function $f(z)$ is a Gaussian distribution with a mean value and Gaussian error. Usually speaking, the key of the GP is a covariance function $k(z,\tilde{z})$ depending on two hyper-parameters $l$ and $\sigma_f$, which describe the coherent scale of the correlation in $x$-direction and typical change in $y$-direction, respectively. In this analysis, we carry out the GP reconstruction by utilizing the available online package GaPP (Gaussian processes in python) \cite{27,28}. Furthermore, we adopt the Mat\'{e}rn ($\nu=9/2$) covariance function implement the reconstruction, which has been proved to be a better choice than the squared exponential covariance function from the statistical point of view (see Ref. \cite{29}). The Mat\'{e}rn ($\nu=9/2$) covariance function is written as
\begin{equation}
k(z,\tilde{z})=\sigma_f^2 \mathrm{exp}(-\frac{3|z-\tilde{z}|}{l})\times[1+\frac{3|z-\tilde{z}|}{l}+\frac{27(z-\tilde{z})^2}{7l^2}+\frac{18|z-\tilde{z}|^3}{7l^3}+\frac{27(z-\tilde{z})^4}{35l^4}]. \label{11}
\end{equation}
Different from the previous works \cite{10,11,28}, we transform directly the LD data from two ADD samples to available normalized comoving distance data by using the formula $D(z)=H_0(1+z)^{-1}D_L(z)$. The EoS of DE can be expressed as \cite{28}
\begin{equation}
\omega(z)=\frac{2(1+z)(1+\Omega_{k})D''-[(1+z)^2\Omega_{k}D'^2-3(1+\Omega_{k}D^2)+2(1+z)\Omega_{k}DD']D'}{3\{(1+z)^2[\Omega_{k}+(1+z)\Omega_{m}]D'^2-(1+\Omega_{k}D^2)\}D'}, \label{2}
\end{equation}
where the primes denote the derivatives with respect to $z$. In the reconstruction processes, we set the initial conditions $D(z=0)=0$ and $D'(z=0)=1$.
Since the constrained values of cosmic curvature from two ADD samples are very close to zero, for simplicity, we fix $\Omega_{k}=0$. Meanwhile, we also use $\Omega_{m}=0.308\pm0.012$ \cite{30} from the recent measurement by the Planck Collaboration. Furthermore, since the value of $H_0$ affects the transformed data \cite{10}, we consider the following two cases for both ADD samples, i.e., the local measurement $H_0=73.24\pm1.74$ km s$^{-1}$ Mpc$^{-1}$ and the global measurement $H_0=66.93\pm0.62$ km s$^{-1}$ Mpc$^{-1}$. It is also interesting to study the effects of constrained $H_0$ values on the reconstructed EoS of DE. In particular, we choose $H_0=70.1\pm0.8$ ($\Lambda$CDM) km s$^{-1}$ Mpc$^{-1}$ for the first ADD sample, and $H_0=74.6\pm2.1$ ($\Lambda$CDM) and $79.0^{+4.7}_{-5.6}$ ($\omega$CDM) km s$^{-1}$ Mpc$^{-1}$ for the second one, respectively. In addition, to make a simple comparison with our previous works \cite{10,11}, we also reconstruct the EoS of DE by adding a data combination of 580 SNe Ia \cite{31}, 30 cosmic chronometers \cite{32}, Planck-2015 shift parameter \cite{33} and 156 HII galaxy measurements \cite{34,35,36,37,38,39,40} into our ADD analysis. The reconstructed results are shown in Figs. \ref{f3}-\ref{f7}.

For the first sample, one can easily find that, when $H_0=73.24\pm1.74$ and $66.93\pm0.62$ km s$^{-1}$ Mpc$^{-1}$, the reconstructions of $D(z), D'(z)$ and $D''(z)$ are very consistent with the prediction of the $\Lambda$CDM model at the $2\sigma$ CL (see Fig. \ref{f3}). Furthermore, from Fig. \ref{f4}, one can find that the reconstructed EoS of DE is also compatible with the $\Lambda$CDM model at the $2\sigma$ CL, when considering R16 and P16's measurements and the constrained value $H_0=70.1\pm0.8$ from the $\Lambda$CDM model. This indicates that there is no evidence of the DDE from the first ADD sample.

For the second example, when $H_0=73.24\pm1.74$ and $66.93\pm0.62$ km s$^{-1}$ Mpc$^{-1}$, one can find that the reconstructions of $D(z), D'(z)$ and $D''(z)$ are just consistent with the $\Lambda$CDM model in the relatively low redshift range at the $2\sigma$ CL (see Fig. \ref{f5}). Meanwhile, interestingly, the reconstructed EoS of DE exhibits a phantom-crossing behavior in the relatively low redshift range over the $2\sigma$ CL, especially for the case of $H_0=66.93\pm0.62$ km s$^{-1}$ Mpc$^{-1}$ (see Fig. \ref{f6}), which gives a hint that the universe may be actually dominated by the DDE from galaxy cluster scales.

By using a combined constraint from ADD, SNe Ia, cosmic chronometers, Planck-2015 shift parameter and HII galaxy measurements, we obtain the following three conclusions: (i) the constraints on the EoS of DE improve significantly and give DDE behaviors over the $2\sigma$ CL (see Figs. \ref{f4}, \ref{f6} and \ref{f7}); (ii) two ADD samples have no evident effects on the constraint on the EoS of DE, i.e., the reconstructions by added four cosmological probes are very stable; (iii) the constraints on the EoS of DE are more tighter than our previous works \cite{10,11}, where we use either a combination of SNe Ia, cosmic chronometers and Planck-2015 shift parameter or only HII galaxy data.

\section{Discussions and conclusions}
Understanding the cosmic acceleration and current $H_0$ tension is two important and urgent tasks in modern cosmology. In this study, we make attempts to explore these two issues using only galaxy cluster data. Based on an assumption that the CDDR is valid ($\eta=1$), we transform the observed data of two ADD samples to effective LD data using the so-called CDDR.

At first, we place constraints on four different cosmological models utilizing the usual Bayesian analysis. For the first ADD sample, we find that the constrained values of $H_0$ from four models are more consistent with R16's local measurement at the $1\sigma$ CL than P15's global one. Using the AIC and BIC, we find that the $\Lambda$CDM model is preferred (see Table. \ref{t2}). Doing the same for the second ADD sample, we find that, although higher $H_0$ values than the first sample are obtained, they are also more consistent with the R16's result at the $1\sigma$ CL and the $\Lambda$CDM model is still preferred with positive evidence (see Table. \ref{t3}). Meanwhile, there is no evidence to show the preference between the left three models for two samples. In addition, we find that a flat universe is still preferred for the first ADD sample. However, this is not the case for the second one, i.e., a positive cosmic curvature corresponding to negatively-curved 3-geometries is mildly supported by 38 ADD data points at the 1$\sigma$ CL. For both ADD samples, the constrained EoS of DE in the $\omega$CDM model is consistent with $\omega=-1$ ($\Lambda$CDM) at the 1$\sigma$ CL, and there is no indication of momentum transfer in the dark sector of the universe from constrained values of $\epsilon$ in the DV model, which are compatible with zero at the 1$\sigma$ CL.

Furthermore, we employ the model-independent GP method to investigate the evolution of the EoS of DE from galaxy cluster data. We find that, for the first sample, when $H_0=73.24\pm1.74$ and $66.93\pm0.62$ km s$^{-1}$ Mpc$^{-1}$, the reconstructions of $D(z), D'(z)$ and $D''(z)$ are very consistent with the prediction of the $\Lambda$CDM model at the $2\sigma$ CL, and that the reconstructed EoS of DE is also consistent with the $\Lambda$CDM model at the $2\sigma$ CL. This implies that there is no evidence of the DDE from the first ADD sample. However, for the second one, when $H_0=73.24\pm1.74$ and $66.93\pm0.62$ km s$^{-1}$ Mpc$^{-1}$, we conclude that the reconstructions of $D(z), D'(z)$ and $D''(z)$ are just consistent with the $\Lambda$CDM model in the relatively low redshift range at the $2\sigma$ CL, and that the reconstructed EoS of DE exhibits a phantom-crossing behavior in the relatively low redshift range over the $2\sigma$ CL, especially for the case of $H_0=66.93\pm0.62$ km s$^{-1}$ Mpc$^{-1}$ (see Fig. \ref{f6}), which gives a hint that the universe may be actually dominated by the DDE from galaxy cluster scales.

It is noteworthy that we implement the reconstructions by assuming $\Omega_{k}=0$ throughout this work, since the constrained values of $\Omega_{k}$ from two ADD samples are very close to zero. Actually, a very small deviation from $\Omega_{k}=0$ may have a large impact on the reconstruction of EoS of DE, especially at high redshifts. This interesting issue will be explored in the forthcoming work. 

In order to make a comparison with our previous works \cite{10,11}, we also reconstruct the EoS of DE using a combination of ADD, SNe Ia, cosmic chronometers, Planck-2015 shift parameter and HII galaxy measurements. Interestingly, we find that this data combination can improve significantly the constraint on the EoS of DE and gives a DDE behavior over the $2\sigma$ CL. Different data combinations effects on the cosmological GP reconstructions are worth being discussed further.

Throughout this analysis, since the values of $H_0$ affect the reconstruction processes, we use $H_0=73.24\pm1.74$ and $66.93\pm0.62$ km s$^{-1}$ Mpc$^{-1}$ in light of the current $H_0$ tension to test the stability of our results. From Figs. \ref{f3}-\ref{f6}, we conclude that the constraints on the EoS of DE are very stable and insensitive to the choice of $H_0$ for both ADD samples. In addition, one can also find that the constraints on the EoS of DE in Figs. \ref{f3}-\ref{f6} appear to diverge when redshift $z$ approaches 1. This can be ascribed to a lack of high redshift data for both ADD samples.

Since the constraints on different model parameters and the reconstructed EoS of DE are different for the two ADD samples, it is worth discussing the differences between both samples. In Figs. \ref{f3} and \ref{f5}, one can easily find that the second sample has larger error bars than the first one, although it has more data points. Hence, there is no overlap between two samples which cover the similar redshift range. Different morphologies, i.e., an elliptical $\beta$ model for the first sample and a generalized $\beta$ spherical model for the second one, also affect the properties of cluster samples. Notice that all cluster mass measurements derived from X-ray and dynamical observations are sensitive to the assumptions about cluster symmetry. In addition, reanalyzing their measurements with a simple isothermal model of the hot plasma without the hydrostatic equilibrium assumption, the authors in Ref. \cite{15} also found that their X-ray/SZ method used to determine the cosmic distance scale is largely insensitive to the details of the hot plasma modeling. Moreover, different methods, systematics and observational techniques applied by two groups may be other important reasons which lead to our different conclusions. The differences between both samples is worthy of being investigated in a forthcoming work. We expect more and more high-quality data with improved probing techniques and methods can be used to explore the unknown dark sector of the universe.
\section{acknowledgements}
This work is partly supported by the National Natural Science Foundation of China. We warmly thank the anonymous referees for improving our analysis. D. Wang thanks B. Ratra and S. D. Odintsov for helpful discussions on gravity and cosmology, and Y. Sun for insightful communications.

\end{document}